\documentclass{aa}
\usepackage{aabib99}
\usepackage{psfig,amssym}
\begin{document}
\thesaurus{03  (11.01.2; 11.10.1; 11.14.1; 11.17.3; 13.18.1)}

\title{On the viewing angle to broad-lined radio galaxies}
\titlerunning{Viewing angle to BLRG.}
\author{Dennett-Thorpe J. \and Barthel P.~D. \and van Bemmel I.~M.}
\institute{Kapteyn Institute, University of Groningen, NL-9700AV
Groningen, Netherlands}
\offprints{J. Dennett-Thorpe}
\mail{jdt@astro.rug.nl, pdb@astro.rug.nl} 
\date{Received 8 September 2000/Accepted}
\maketitle

\begin{abstract} We address the nature of broad-lined
radio galaxies, in particular their radio axis orientation, using new,
matched resolution, dual frequency radio observations of a sample of
twelve nearby broad-lined extragalactic 3C objects.  Radio spectral
index and depolarisation asymmetries indicate that these objects have
a preferred orientation with respect to the observer. In addition, the
spectral asymmetries are suggestive of lower Doppler factors in the
broad-lined radio galaxies when compared to 3C quasars. This is in
agreement with their optical properties, and leads to the conclusion
that some objects are lower powered versions---at similar lines of
sight---of the more distant quasars, whereas others are at larger
angles to the line of sight.

\end{abstract}

\section{Introduction} 

Under most current paradigms of active galactic nuclei, the broad
emission lines seen in the optical spectra of such objects are due to
emission originating in regions close to a supermassive black hole.
Orientation-dependent unification schemes (e.g. Scheuer 1987, Barthel
1989) posit that at large viewing angles this broad-line region is
obscured by optically thick material (a putative dusty torus).  Within
this scheme, radio-loud quasars have been proposed to represent
favorably oriented narrow-lined radio galaxies.
\nocite{sch87-contrib,bar89}

Broad-lined radio galaxies (BLRGs) have both a visible host galaxy
(hence their identification on optical images as galaxies) and broad
lines in their spectra.  On account of these broad emission lines, BLRGs
can be considered to be aligned to the observer at similar angles to the
line of sight as the quasar population.  This assumption is often made
in statistical investigations of radio galaxies and quasars which are concerned
with orientation, obscuration and relativistic motion.  There is indeed evidence
that some BLRGs are at comparable angles to the lines of sight to
quasars (e.g.  superluminal motion in some BLRGs \cite{ale96}) and that,
statistically, they are closer to the line of sight than their
narrow-lined counterparts (Hardcastle et al. 1998; Gilbert \& Riley
1999).  \nocite{1998MNRAS.296..445H,1999MNRAS.309..681G} From an
analysis of the radio properties of BLRGs in the 3CR sample, Hardcastle
et al.  \cite*{1998MNRAS.296..445H} argue that BLRGs are simply
low-luminosity counterparts of quasars.  In this picture, the objects
are classified as BLRGs because of the combined effects of the lower
luminosity of the AGN and the proximity to the host enabling the
starlight to be clearly detected. 

However, there are two lines of evidence arguing against the simple
picture that all broad-lined radio galaxies are nearby, low-luminosity
quasars.  Tadhunter et al.  \cite*{1998MNRAS.298.1035T} detect BLRGs out
to the redshift limit of their sample, thus overlapping with the quasars
in redshift and power.  This suggests a considerable range in the
intrinsic luminosities of the broad-line AGN for a given redshift or
radio power.  The authors favour this explanation over the possibility
that the BLRGs are simply partially obscured quasars, because of the
large scatter seen in the correlation between radio and [OIII]
luminosities.  Secondly, BLRGs have been shown to be unusual in their
FIR spectral energy distribution.  Many BLRGs show a short wavelength
FIR excess seen neither in NLRGs nor in quasars, which is presumably due
to hot dust (Hes et al.  1995).  In fact, the first BLRG studied by IRAS
already displayed this hot dust component (Miley et al.  1984).  Further
work is underway to investigate how general this property is, using the
ISO satellite (van Bem\-mel et al., in prep.). 
\nocite{1984ApJ...278L..79M,hes95}

Not all the broad-line radiation may be reaching us directly.  Combining
results from optical spectropolarimetry with information on dust-induced
reddening, Cohen et al.  (1999) suggest a continuous transition with
increasing aspect angle from unobscured quasars to reddened BLRG to
NLRG.  This argues that the obscuring torus does not have a distinct
edge (see also Baker 1997).  Cohen et al.  model the optical AGN
radiation of BLRGs with contributions from both dust-reddened and
dust-scattered quasar light.  Corbett et al.  (1998) find that
broad-line emission may be scattered both from the inner wall of the
postulated dust torus and from its polar regions by electrons or dust
clouds distributed along the radio jet.

It therefore seems likely that both the luminosity of the AGN itself
and its orientation play a role in whether an object is classified as
a BLRG or a quasar. In this paper we address the relative importance
of these effects using radio and other data. Determining the
orientation of the BLRG axis will be essential to discriminate between
the possibilities and determine the location and nature of the hot
dust.  \nocite{1998MNRAS.296..721C}
 
In addition to the brightness asymmetries of the jets and hotspots,
which are crude indicators of orientation (see e.g.  Hardcastle et al. 
1998), there are two other orientation indicators in the radio emission:
depolarisation and hotspot spectral index asymmetries.  Depolarisation
asymmetries related to orientation of powerful classical double radio
sources was first reported by Laing (1988) and Garrington et al. 
(1988).  Using a sample of quasars and radio galaxies with one detected
jet, these studies show that the jet side is generally less
depolarised.  This is most easily understood as an orientation effect,
in which the jet is pointing towards us and the counterjet side is seen
through a greater depth of depolarising material associated with the
host galaxy. 

Spectral asymmetries related to jet-sidedness in powerful radio sources
with detected one-sided jets were first reported by Garrington et al. 
(1991).  The observations in that paper were insufficient to ascertain
if the asymmetry was due to the hotspot material, or the extended lobe
material: a situation complicated by the detection of spectral
asymmetries in the extended material in another sample \cite{liu91b}. 
Dennett-Thorpe et al.  (1997; hereafter D97) addressed this issue and
showed that the spectral asymmetries could be separated into two
components: one in the lobe material related to the confinement of the
radio lobe and a second, of interest here, in the hotspots due to
relativistic effects. The spectral asymmetry of the hotspots has also
been confirmed by  Ishwara-Chandra \& Saikia (2000).

In this paper we present results of dual frequency radio observations of
BLRGs and quasars taken for the purpose of investigating
their orientation by use of the spectral and depolarisation asymmetries. 
We present the observations and data analysis (Sect.~\ref{sect:data})
before presenting the results, also combined with the data from D97 and
D99 (Sect.~\ref{sect:results}).  In Sect.~\ref{sect:interp} we present
modelling of the expected spectral asymmetries, also in the presence of
lobe contamination, and analyse the observed asymmetries related to the
object classification (BLRGs and quasars).  We finally consider other
evidence on the viewing angles to the individual broad-lined objects
considered in this paper, using data from the literature in
conjunction with our results. 

Throughout this paper we use a Friedmann cosmology with H$_0$=65
km\,s$^{-1}$Mpc$^{-1}$ and q$_0$=0.5. We use the convention S$\propto
\nu^{-\alpha}$.

\section{Sample selection, Observations and data reduction}
\label{sect:data}

The sample was chosen to include all broad-lined 3C objects in the
redshift interval 0.15$<z<$0.36, which exceed 20\,arcsec in angular
size.  The size criteria ensures that all sources will be resolved by
the observations.  The upper limit of the redshift range was chosen to
include all objects commonly considered as broad-lined radio galaxies,
whilst the lower limit corresponds to the upper redshift limit of the 3C
radio galaxy sample of Dennett-Thorpe et al.  (1999; hereafter D99). 
That investigation involved a similar analysis, and contained two
broad-lined radio galaxies (3C\,382 and 3C\,390.3), which are also
included in part of the present analysis. 

The previous detection of a jet was not a criterion for inclusion in the
sample.  We observed both known quasars and objects catalogued as BLRG. 
We excluded the blazar 3C\,273 from our sample.  The quasar 3C\,249.1,
which falls in the redshift range of our sample, was not observed in
this programme, but was observed by D97, and is included in our
analysis.  The properties of the sample are summarised in
Table~\ref{tab:sources}.  In the first three columns we give the 3C and
IAU names, and the identification as narrow-line object (N); broad-line
galaxy (B); or quasar (Q) (see Sect.~\ref{sect:class}).  The next
three columns give fundamental source parameters: redshift,
monochromatic radio power at 1.4\,GHz and largest angular size.  The
rotation measure and side of any detected jet are given in the final
columns. 

The observations were taken with the NRAO VLA at 1.4\,GHz in B array
on 3 Oct 1999 and at 4.9\,GHz in C array on 1 Dec 1998.  The arrays
were chosen to deliver matched {\sl uv} coverage at the two
frequencies, which is crucial for spectral work.  All reductions were
done in {\sc aips}.  The absolute flux density and position angle
calibration was done using 3C286.  Other nearby amplitude calibrators
were interspersed throughout the run.  In order to maximise {\sl uv}
coverage, each source was observed several times (typically 6, with
some restrictions due to scheduling constraints) over a period of
several hours.  Typical observational parameters are listed in
Table~\ref{tab:obs}.  Baseline solutions were applied in the
conventional manner by observing the sources 0410+769 and 1400+621.

Due to telescope failure 3C\,286 was not observed in the 1.4\,GHz
observations, but was observed two days later, together with sources
1400+621 and 3C\,287.1.  In this way the absolute flux density was tied
to the non-variable CSO \\1400+621, and the stability of the
polarisation solutions over the intervening two days could be checked by
imaging 3C\,287.1.  Comparison of the observations showed the position
angle and percentage polarisation to be the same within $<$3\% for both
runs, indicating that, as expected, the telescopes were stable over this
time interval. 

Data for all sources were phase self-calibrated to convergence, and the
final uniformly-weighted images were produced by cleaning to the noise
with {\sc aips} task {\sc imagr}.  Particular care was taken to ensure
the resultant images were cleaned sufficiently deep: spectral index work
is particularly sensitive to the erroneous flux densities that can arise
by restoring the clean conponents (mJy per clean beam) onto an
insufficiently cleaned image (with flux density measured in mJy per
dirty beam).  The polarisation images were corrected for Ricean bias
using the conventional {\sc aips} tasks for this purpose. 

The 5\,GHz images are presented in Appendix A.  The experiment was
designed to yield identical resolution at both frequencies, and has
higher sensitivity at 5\,GHz.  As a
result, the 1.4\,GHz images are virtually identical and are not shown
here.

\begin{table*}
\caption{Source properties}
\label{tab:sources}
\begin{tabular}{lllllrll}
\hline
\hline
Name     & IAU name  &ID&redshift & log(P$_{\rm 1.4GHz}$)& LAS &RM &jet-side \\
         &           &  &  & (W Hz$^{-1}$)      & (arcsec) & (rad m$^{-2})$ &\\
\hline
3C\,17   &0035$-$024&B & 0.220  & 26.11   &    45  & $+10 \pm$3 &SE\\
3C\,33.1 &0106+729&B & 0.181  & 25.62   &   240  & $-15 \pm$3 &S \\
3C\,61.1 &0210+860&N & 0.184  & 25.91   &   185  &    --      &--\\
3C\,93   &0340+048&B & 0.357  & 26.40   &    27  & $ +9 \pm$2 &N?\\
3C\,109  &0410+110&B & 0.306  & 26.33   &   100  & $-16 \pm$2 &S\\
3C\,206  &0837$-$120&Q & 0.197  & 25.43   &   169  &$-105 \pm$7 &E \\
3C\,219  &0917+458&B & 0.174  & 25.99   &   180  & $-19 \pm$2 &S\\
3C\,234  &0958+290&B & 0.185  & 25.86   &   120  & $+42 \pm$1 &NE\\
3C\,246  &1048$-$090&Q & 0.344  & 26.10   &    83  & $ +0 \pm$3 &--\\
3C\,287.1&1330+023&B & 0.216  & 25.76   &   110  & $ +1 \pm$2 &W \\
3C\,323.1&1545+210&Q & 0.264  & 25.94   &    68  & $+14 \pm$0.4&S\\
3C\,332  &1615+324&B & 0.152  & 25.34   &    90  & $ +2 \pm$1 &S \\
3C\,381  &1832+474&B & 0.160  & 25.55   &    75  & $+25 \pm$1 &N?\\
\hline
\end{tabular}

\underline{References}\\ 
{\it Redshifts:} 3C\,17: Schmidt \cite*{1965ApJ...141....1S};
3C\,33.1: Chambers et al. \cite*{1996ApJS..106..247C}; 3C\,61.1:
Lawrence et al.\cite*{1996ApJS..107..541L}; 3C\,93:
Hewitt \& Burbidge \cite*{hewitt}; 3C\,109,219,234,332,381: Hewitt \&
Burbidge \cite*{1991ApJS...75..297H}; 3C\,206: Ellingson et
al. \cite*{1989AJ.....97.1539E}; 3C\,246: Schmidt \& Green
\cite*{1983ApJ...269..352S}; 3C\,287.1: Dunlop et
al. \cite*{1989MNRAS.238.1171D}; 3C\,323.1: Marziani et
al. \cite*{1996ApJS..104...37M}\\
{\it Radio power:} Computed from White \& Becker (1992), except
3C\,206 \& 3C\,246 for which we used PKSCAT90 (Wright \& Otrupcek
1990),\nocite{pkscat} and 3C\,61.1 (K{\"u}hr et
al. 1981).\nocite{1981A&AS...45..367K,1992ApJS...79..331W}\\
{\it RM:} Simard-Normandin et al. \cite*{1981ApJS...45...97S} \\
{\it Radio maps:} 3C\,17: Morganti et
al. \cite*{1999A&AS..140..355M};
3C\,33.1: http://www.jb.man.ac.uk/atlas/; 3C\,61.1: Leahy \& Perley
\cite*{1991AJ....102..537L}; 3C\,93: Harvanek \& Hardcastle
\cite*{1998ApJS..119...25H}; 3C\,109: Giovannini et
al. \cite*{1994ApJ...435..116G}; 3C\,219: Perley et
al. \cite*{1980AJ.....85..499P}; Clarke et
al. \cite*{1992ApJ...385..173C}; 3C\,234: Burns et
al. \cite*{1984ApJ...283..515B}; Hardcastle et
al. \cite*{1997MNRAS.288..859H}; 3C\,287.1: Antonucci
\cite*{1985ApJS...59..499A}; 3C\,332: Fanti et
al. \cite*{1987A&AS...69...57F}; 3C\,381: Leahy \& Perley
\cite*{1991AJ....102..537L}; Hardcastle et
al. \cite*{1997MNRAS.288..859H}; 3C\,323.1: Bogers et al.
\cite*{1994A&AS..105...91B};
\end{table*}

\begin{table*}
\caption{Observation parameters}
\label{tab:obs}
\begin{tabular}{lllllll}
\hline
\hline
date &\multicolumn{2}{c}{frequency} & bandwidth &typical scans &integration & r.m.s. noise\\
     &\multicolumn{2}{c}{(GHz)}     & (Mhz)     &per source  & time (mins) & (mJy) \\
\hline
3 Oct 1998 & 1.4650& 1.6650 &  25.0     &  6    & 45 & 0.4 \\
1 Dec 1998 & 4.8351& 4.8851 &  50.0     &  6    & 45 & 0.1 \\
\hline
\end{tabular}
\end{table*}

\subsection{Classifications}
\label{sect:class}

The classification of an object as a quasar or a radio galaxy has been
done in the past by a number of different workers.  We therefore checked
several characteristics ourselves to ensure homogeneity of definition
across the sample.  Firstly we checked the Digitised Sky Survey (of the
second generation POSS plates) for evidence of the host galaxy.  We
concluded 3C\,246 and 3C\,323.1 were clearly stellar, 3C\,93, 3C\,234,
3C\,287.1 and 3C\,109 had some evidence for host galaxy or surrounding
nebulosity, whilst on the remainder the host galaxy was clearly seen. 
The E--O colours were compared using the APM (which had no information
for 3C\,33.1, 3C\,61.1, 3C\,206), as bluer colours indicate a more
dominant AGN contribution.  3C\,246 and 3C\,323.1 (both PG quasars) were
the bluest objects, followed by 3C\,287.1, 3C\,234, 3C\,93 and 3C\,17. 

Spectral confirmation of the objects' broad emission lines for most
objects came from spectra published in Eracleous \& Halpern
\cite*{1994ApJS...90....1E} (see note on 3C\,61.1 below).  For 3C\,33.1,
3C\,219 and 3C\,381 other references were used (referenced in
Appendix~B). 

Notes on individual sources can be found in Appendix B. Here we bring
attention to the fact that 3C\,61.1 has been previously misidentified
as a broad-line object.  3C\,381, although also reported to have broad
lines, has no published spectrum confirming this.  3C\,234 is a
well-known broad-line object: it has broad wings on the lines in its
total intensity spectrum, but these are much more prominent in
polarised light, indicating that the broad-line region is partially
obscured.  3C\,93 and 3C\,109 have been subject to extensive studies
which suggest their nuclear regions are reddened.  In summary, for the
analysis of the radio asymmetries, we consider 3C\,206, 3C\,246 and
3C\,323.1 as quasars, and the other objects (except 3C\,61.1) as
BLRGs.  We return to these definitions later.

\section{Empirical results}
\label{sect:results}

As not all the objects have detected jets, it is not possible to analyse
all sources in terms of jet/counterjet side directly.  We therefore
allocate sides A and B to the structure, as indicated in column 2 of
Table~\ref{tab:results}.  Where possible this has been associated with
the jet side, such that side A corresponds to the jet side.  The sources
in which the jet-side is not known, or ambiguous, are: 3C\,61.1, 3C\,93,
3C\,246 and 3C\,381.  In three of these cases we have assigned a
jet-side based on some hints of jet-like features in our maps, or those
referenced in Table~\ref{tab:sources}, whilst for 3C\,246 the assignment
was arbitrary. 

\begin{table*}
\caption{Spectral and (de)polarisation properties.}
\label{tab:results}
\begin{tabular}{llrrrrrrrrrrr}
\hline
\hline
source & side A&$\alpha_A$ & $\alpha_B$ &$\alpha_A^{\rm hs}$ &
$\alpha_B^{\rm hs}$& \multicolumn{2}{c}{$(\alpha_B^{\rm hs} -
\alpha_A^{\rm hs})$} &\%P$_A^5$ & \%P$_B^5$ &
DP$_A$& DP$_B$ &DP$_A$/DP$_B$\\
\hline
3C\,17   & SE & 0.81& 0.89& 0.85& 1.27& 0.42& --   & 06.5 & 07.9 & 0.355& 0.329& 1.08 \\
3C\,33.1 & S  & 0.94& 0.94& 0.77& 0.80& 0.03&  0.03& 19.6 & 14.1 & 0.922& 0.913& 1.01 \\
3C\,61.1 & N  & 1.03& 0.98& 0.94& 0.96& 0.03&  0.03& 24.2 & 16.1 & 0.269& 0.297& 0.91 \\%
3C\,93   & N  & 0.88& 0.94& 0.81& 0.92& 0.11&  0.09& 06.2 & 14.7 & 0.356& 0.326& 1.09 \\%
3C\,109  & S  & 0.94& 0.94& 0.87& 0.94& 0.07&  0.08& 12.1 & 12.3 & 0.314& 0.308& 1.02 \\%
3C\,206  & E  & 0.94& 0.93& 0.78& 0.78& 0.00&  0.00& 15.8 & 15.8 & 0.316& 0.317& 1.00 \\%
3C\,219  & S  & 1.02& 1.02& 0.79& 0.83& 0.04&  0.04& 14.7 & 15.8 & 0.292& 0.304& 0.96 \\%
3C\,234  & NE & 1.00& 1.03& 0.96& 1.03& 0.06&  0.06& 15.9 & 12.7 & 0.289& 0.283& 1.02 \\%
3C\,246  & W  & 0.89& 0.92& 0.86& 0.90& 0.03&  0.03& 04.2 & 06.0 & 0.355& 0.315& 1.13 \\%
3C\,287.1& W  & 0.81& 0.84& 0.83& 0.82&-0.01& -0.01& 12.8 & 13.0 & 0.360& 0.346& 1.04 \\%
3C\,323.1& S  & 0.86& 0.86& 0.87& 0.84&-0.02& -0.03& 06.0 & 08.4 & 0.333& 0.333& 1.00 \\%
3C\,332  & S  & 0.86& 0.88& 0.74& 0.77& 0.04&  0.03& 16.3 & 13.4 & 0.338& 0.328& 1.03 \\%
3C\,381  & N  & 0.82& 0.84& 0.77& 0.84& 0.07&  0.08& 07.7 & 16.2 & 0.945& 0.918& 1.03 \\%
\hline
\end{tabular}	  		     
\end{table*}

\subsection{Spectral asymmetries}

Table~\ref{tab:results} summarises the results of the analysis of the
radio asymmetries.  Columns 3 \& 4 give the spectral indices of the
entire radio lobes, excluding the cores, but including the jets and
hotspots.  Columns 5, 6 \& 7 give the spectral indices of the hotspots,
and spectral difference calculated in the manner of D97.  This method
uses the brightest regions of the source and integrates the flux in this
region in the images at both frequencies.  Using this method we are able
to compare directly with the results obtained in that paper.  In column
8 we present the spectral index difference calculated in the more
conventional way of using a two-component fit to the maximum: a single
Gaussian component and a zero-level and slope ({\sc aips} task {\sc
imfit}), where the peak of the Gaussian component is used to calculate
the spectral index of the hotspot.  We see very good agreement between
the different methods of calculation (i.e. columns 7 \& 8). 

For 3C\,17 there is no good fit to such a Gaussian model at this
resolution.  The strong jet which is apparent in the higher resolution
images (Morganti et al. 1999) is unresolved in this object---the only
source for which this is the case.  Therefore the (presumably) flat
spectrum jet emission contributes substantially to the calculated jet
side spectral index, and is undoubtedly the cause of the large spectral
asymmetry in this object. 

3C\,206 has a bright compact feature far from the end of the radio
lobe on the counterjet-side.  We have taken this to be the hotspot,
because sites of dramatic jet disruption and jet `termination'
are difficult to distinguish both theoretically and observationally.

Errors on spectral indices are dominated by the absolute flux density
calibration, which we take as 3\%, given that good solutions were
obtained for 3C\,286.  However, in comparison of spectral index
differences in parts of a single source, this calibration uncertainty
cancels out, leaving $<$1\% error due to noise alone.  Thus the spectral
index {\it asymmetry} in the hotspot regions has an error $<$0.01 from
this origin. 

\begin{figure}
\psfig{file=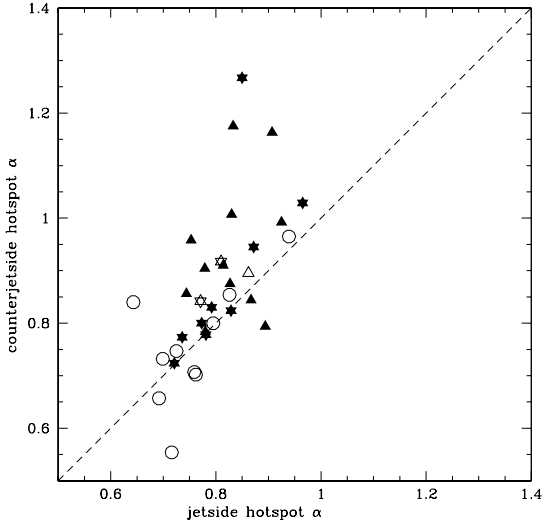,height=7.5cm}
\caption{Spectral indices of hotspots of BLRGs (stars), quasars
(triangles) and narrow-line radio galaxies (open circles) in this
paper combined with those in D97 and D99. BLRGs and the quasar with uncertain
jet-sidedness are shown as open stars and triangle.}
\label{fig:spix}
\end{figure}

The results are presented in Fig.~\ref{fig:spix}, which also includes
the results of the 3CR radio galaxies and quasars from D97 and D99.  Two
broad-line radio galaxies, 3C\,382 and 3C\,390.3 were studied in D99:
interestingly these had shown no detectable spectral asymmetry in the
hotspot regions. 

It can be seen that the narrow-line radio galaxies (open circles) show
asymmetries in both senses: they are scattered evenly both above and
below the dashed line, indicating that in some sources the jet side
hotspot has a flatter spectrum, whilst in others its spectrum is
steeper.  The quasars and the broad-line radio galaxies on the other
hand show a strong bias towards having a flatter jet side spectrum in
the hotspot region: they fall above the dotted line.  Considering only
the broad lined objects presented in this paper, with detected jets and
spectral asymmetries, 6/8 have detectably flatter jet side hotspots. 
The probability that at least this number have flatter jet side hotspots
by chance is quite substantial (14\%) but is strongly affected by small
number statistics.  Considering the combined results of this paper
(objects with detected jets only), D97 and D99, we find that 9/11
quasars and 6/7 BLRGs have flatter jet side hotspot spectra.  There are,
in addition, 1 quasar and 2 BLRGs with the same counterjet and jet side
spectral indices.  The probability that at least 15/21 sources show
flatter spectrum jetside hotspots by chance is 4\%. 

\begin{figure}
\psfig{file=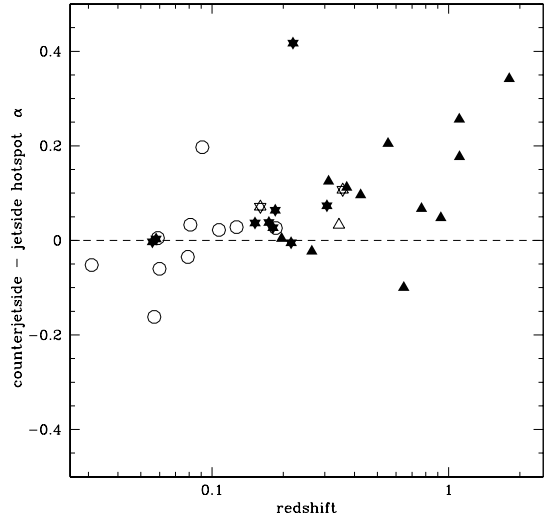,height=7.5cm}
\caption{Spectral asymmetries in hotspots of BLRGs (stars), quasars
(triangles) and narrow-line radio galaxies (circles), (as
Fig.~\ref{fig:spix}) as a function of redshift.}
\label{fig:spix-z}
\end{figure}

There is also a trend seen in Fig.~\ref{fig:spix} amongst the BLRGs and
quasars that the magnitude of the asymmetry increases as the (mean)
hotspot spectrum steepens.  Fig.~\ref{fig:spix-z} shows the spectral
asymmetry as a function of source redshift.  If one excludes the
insufficiently resolved source 3C\,17 ($z$=0.22, $\Delta\alpha$=0.4),
there is also a trend of increasing magnitude of asymmetry with redshift
for the BLRGs and quasars. 

\subsection{Depolarisation asymmetries}

We calculate the percentage polarisation, \%P, of a radio lobe by
integrating the flux density in the Ricean corrected polarised
intensity image, above 5$\sigma$ on the total intensity image (columns
9 \& 10 of Table~\ref{tab:results}).  This is not the same as a single
dish measurement, but has the advantage of allowing for changes in
polarisation position angle across the source, as far as this has been
resolved, without weighting towards the highly polarised boundary
regions.  The depolarisation parameter, DP, is calculated as
\%P$_{\rm 1.4}$/\%P$_{\rm 5}$ (columns 11 \& 12).  The core has
been excluded from these measurements, but not the jet.  If side B
(usually the counterjet side) is more heavily depolarised, the final
column is $>$ 1.

Errors in DP calculated from the noise on the maps alone are small,
typically $\lesssim$1\%.  Absolute flux density calibration will not
affect this quantity, but the polarisation calibration will.  From
inspection of the solutions obtained during the calibration procedure,
we conclude that the fractional polarised flux densities are accurate to
within 3\%.  We therefore take the errors on the depolarisation
parameter to be $\sim$3\%. 

From inspection of Table~\ref{tab:results}, we see that 10 of the
broad-lined objects (thus, excluding 3C\,61.1) show an asymmetric
depolarisation.  Of these, 9 are more heavily depolarised on `side B'. 
Taking the objects with known jet-sidedness only, we see that 6/7
sources have more depolarised counter-jet sides, in good agreement with
the findings of Laing, Garrington et al.  The probability that at least
this number show this asymmetry by chance is 6\%.  The one source
(3C\,219) which shows depolarisation in the other sense, has a flatter
jet side hotspot (i.e.  does not have a spectral asymmetry counter to
the expected sense).  The NLRG 3C\,61.1 has an ambiguous jet detection,
but the side with the flattest hotspot and the least depolarisation are
different, so cannot both indicate the jet side.  This is expected for a
source in the plane of the sky, as environmental effects are known to
cause both spectral and polarisation asymmetries (D99 and references therein). 
Of the BLRGs and quasars with ambiguous or no jet detections, all have
the depolarisation and spectral asymmetries in the sense that the least
depolarised side is also the side with the flattest hotspot.  We believe
that, given the strength of the correlations where the jet side is
known, we may use these asymmetries, and their consistency, to indicate
jet side in all three of these sources.  Thus these sources are marked
in the figures which are labelled in terms of jet-sidedness.  We have,
however, clearly distinguished them (with open stars and triangles). 

\begin{figure}
\psfig{file=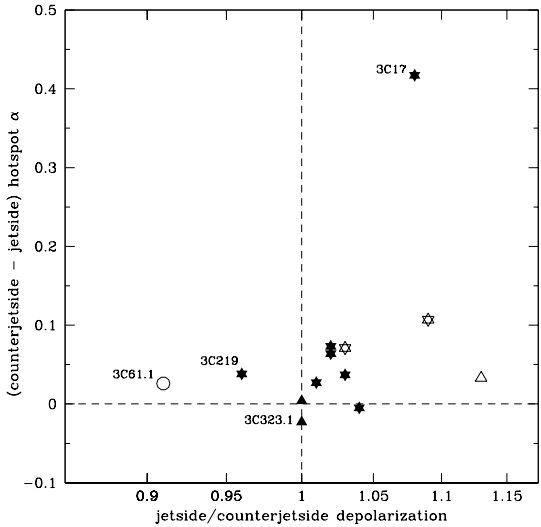,height=7.5cm}
\caption{Depolarisation and spectral asymmetries in hotspots of BLRGs (stars) and quasars
(triangles) and 3C\,61.1(circle) presented in this paper. Although
all of the depolarisation asymmetries detected are marginally
significant,  for the entire sample of BLRGs and quasars both
asymmetries are significant.}
\label{fig:dp-spix}
\end{figure}

Fig.~\ref{fig:dp-spix} shows the correlation of hotspot spectral index
asymmetry and depolarisation asymmetry for the sources observations in
this paper.  Clearly most sources fall in the upper right quadrant
where the hotspot region on the jet side has a flatter spectrum than
on the counterjet side, and the jet side lobe is less depolarised.
Excluding the NLRG 3C\,61.1, we see that only 2/12 sources (3C\,323.1
and 3C\,219) fall well outside this quadrant.

\section{Interpretation}
\label{sect:interp}

\subsection{Hotspot spectral asymmetries}

The results of the preceding section showed that the BLRGs, like the
quasars in D97, have, on average, flatter spectra on the jet side
hotspots.  This can be explained as an effect of relativistic Doppler
boosting.  The simplest explanation \cite{1992MNRAS.256..281T}
attributes this to a higher flux density of flat-spectrum hotspot
material on the jetted side, due to Doppler enhancement of this material
over the quasi-stationary lobe material.  This requires the jet side to
have brighter hotspots at the resolution of the observations, whereas in
fact only 5/12 of the quasars and 4/11 of the BLRGs (with known jet
sidedness) show this effect.  Thus, as argued in D97, the effect must be
due to either Doppler frequency-shifting of curved hotspot spectra, or
to Doppler suppression of the hotspot on the counterjet side. 

\begin{figure*}
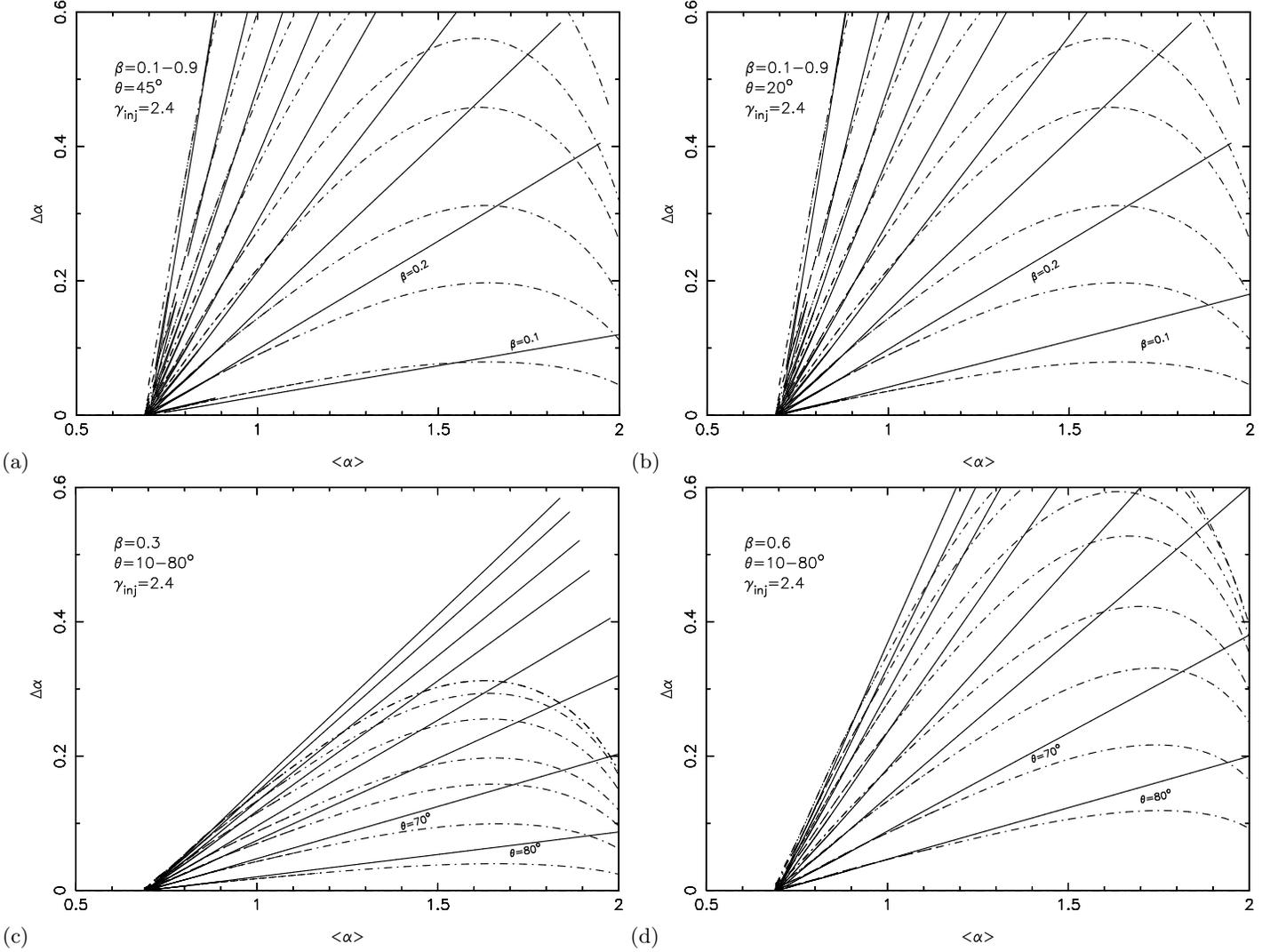

\centerline{(a)\psfig{file=h2439f4a.ps,angle=-90,height=7cm}
(b)\psfig{file=h2439f4b.ps,height=7cm,angle=-90}}
\centerline{(c)\psfig{file=h2439f4c.ps,height=7cm,angle=-90}
(d)\psfig{file=h2439f4d.ps,angle=-90,height=7cm}}
\caption{The spectral asymmetry as a function of average resultant spectral
index for both JP (solid lines) and KP (dotted lines) cases. Results
for motion inclined at (a) 45$^\circ$ and (b) 20$^\circ$ for
$\beta$=0.1 to 0.9 (in steps of 0.1), and for (c)
$\beta =0.3$ and (d) $\beta = 0.6$ at angles to the line of sight
ranging from 10$^\circ$ to 80$^\circ$ (in steps of
10$^\circ$). No underlying continuum is assumed.}
\label{fig:theory}
\end{figure*}

The magnitude of the spectral asymmetry is smaller in the BLRGs than in
the quasars (Fig.~\ref{fig:spix}).  This can be explained, to first
order, as a result of the lower power/redshift of the BLRGs compared to
the quasars.  The interpretation of the spectral asymmetry arising due
to a spectral curvature in the hotspot spectrum predicts such an effect
with redshift, due to the shifting of the emitting frequencies for fixed
observing frequencies.  At higher redshift sources should have both
steeper spectrum hotspots and greater spectral asymmetry due to this
effect, as seen in Fig.~\ref{fig:spix-z}.  This effect was already apparent in
D99.  Analysis conducted in that paper showed that the higher emitted
frequency observed in the higher redshift objects alone is insufficent
to explain the magnitude of the effect, and therefore that part of the
effect must be related to increasing radio power. 

The notion of Doppler shifted curved spectra is well supported by the
fact that the sources with steepest hotspot spectra have the highest
spectral asymmetry (regardless of the power or redshift origin of this
difference).  In order to estimate the Doppler factors that would be
required to produce the observed asymmetries, we assume a standard
(curved) spectrum of the hotspots, and calculate the spectral asymmetry
caused by a relativistic Doppler frequency shift.  In this way the
origin of the steeper spectra in higher power/ redshift objects is
immaterial. 

Fig.~\ref{fig:theory} illustrates both the effects on the observed
spectra of changing the inclination to the line of sight $\theta$, and
the speed of the flow $\beta$.  (It is the Doppler factor {\cal
D}($\beta,\theta$) which uniquely determines the curve, however as we
are more interested in the angle to the line of sight we have presented
the results in this form.) There are two families of curves, both for
tangled magnetic fields, one for pitch-angle isotropised electron
distributions (Jaffe-Perola (JP); solid line) the other for
non-isotropised distributions (Kardashev-Pacholczyk (KP); dashed lines). 
The spectra due to isotropised electrons steepen increasingly with
frequency, so the spectral asymmetry also increases with increasing mean
spectral index (i.e.  frequency).  In the non-isotropised case however,
there is a maximum asymmetry produced (as a function of mean spectral
index, or observing frequency) as the spectrum at high frequency tends
to (4/3)$\alpha_{\rm inj}$+1, where $\alpha_{\rm inj}$ is the low
frequency (injection) spectral index.  Continuous injection models
result in lower asymmetries for a given Doppler factor and mean spectral
index.  We present only the KP/JP cases because these represent the
extremes of possible spectra, and in particular the JP model yields
strong lower limits on the Doppler factors required to produce the
asymmetries.  We have assumed an injection spectral index $\alpha_{\rm
inj}=0.7$.  Other initial electron distributions, with power law
indices, obviously have the effect of shifting the curves along the x-axis. 

\subsection{The effect of contamination by lobe material}

\begin{figure*}
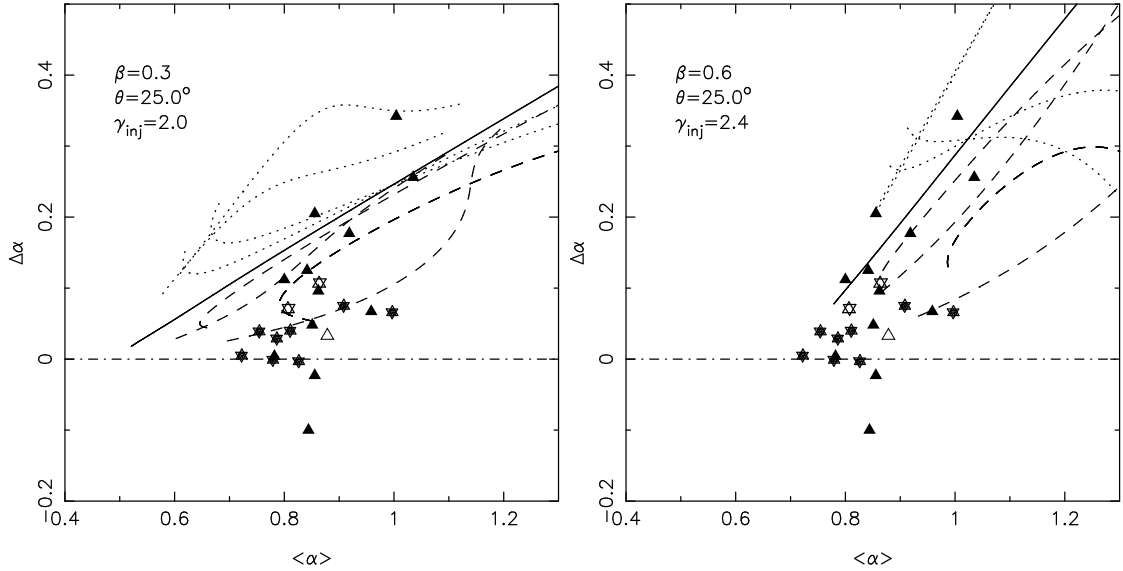

\centerline{\psfig{file=h2439f5a.ps,angle=-90,height=7.5cm}
\psfig{file=h2439f5b.ps,angle=-90,height=7.5cm}}
\caption{The spectral asymmetry as a function of average resultant spectral
index for the JP case, with inclusion of non-boosted lobe
emission. Solid line shows zero unboosted (lobe) emission. The
dotted and dashed lines refer to lobe emission plus Doppler frequency
shifted hotspot emission, where the hotspot emission has also been
Doppler boosted in flux density (dotted) or not (dashed). Lines
tending to $\alpha_{\rm inj}$ (=$(\gamma_{\rm inj}-1)/2$; 0.5 left,
0.7 right) are model A, lines curving away from this are model B.}
\label{fig:underlying}
\end{figure*}

So far we have assumed no contribution from underlying (non-boosted)
material.  The effect of such material---assuming this has a steeper
spectrum than the hotspot material---is to increase the average
observed spectral index of the hotspot region, but the effect on the
spectral index asymmetry is complex, and depends on the assumed
relative flux densities and spectra of the components.  The effects
are shown in Fig.~\ref{fig:underlying} for two models of the lobe
component, described below, for two different particle injection
indices.  In both cases an angle to the line of sight of 25$^\circ$
has been used, and the flow speed has been chosen to ensure the data
points lie near the calculated loci (from inspection of
Fig.~\ref{fig:theory}).

We consider two models of the lobe material.  One model (A) is given
by S$_{\rm lobe}\propto \nu^{-1.5}$, and in the other model (B)
the spectrum of the lobe material has the same spectrum as
the hotspot material, but with the break frequency shifted to
$0.1\nu_b^{\rm hs}$, where $\nu_b^{\rm hs}$ is the intrinsic break frequency
of the hotspots.  For each model take two different ratios of $S_{\rm
hotspot}/S_{\rm lobe}$, satisfying the requirement that the hotspot
remains detectable as such: the surface brightness of the hotspot is
greater than that of the lobe material at a frequency $\sim
0.1\nu_b^{\rm hs}$.

We have plotted the results for the two models, in each case with
(dotted lines) and without (dashed lines) the flux boosting of the
hotspot taken into consideration.  Although flux density boosting must
occur if the hotspots contain relativistically moving material, other
factors have been shown to determine the brightness of the hotspot,
which is critical for the spectral asymmetry. We crudely illustrate
this by neglecting the flux density boosting.

In summary, the effect of the inclusion of the lobe material is complex,
and depends on both the spectral form of the lobe material and the
relative brightness of the hospot to the background material.  The
effect will be to introduce considerable scatter into the observed
hotspot spectral asymmetries.  Therefore, we cannot use the spectral
asymmetries as an accurate line of sight estimator for individual
objects, unless our observations are of sufficiently high resolution to
resolve the hotspots themselves, and thereby ascertain the contribution
from the lobe material. 

\subsection{Comparison of BLRGs and quasars}

\begin{figure*}
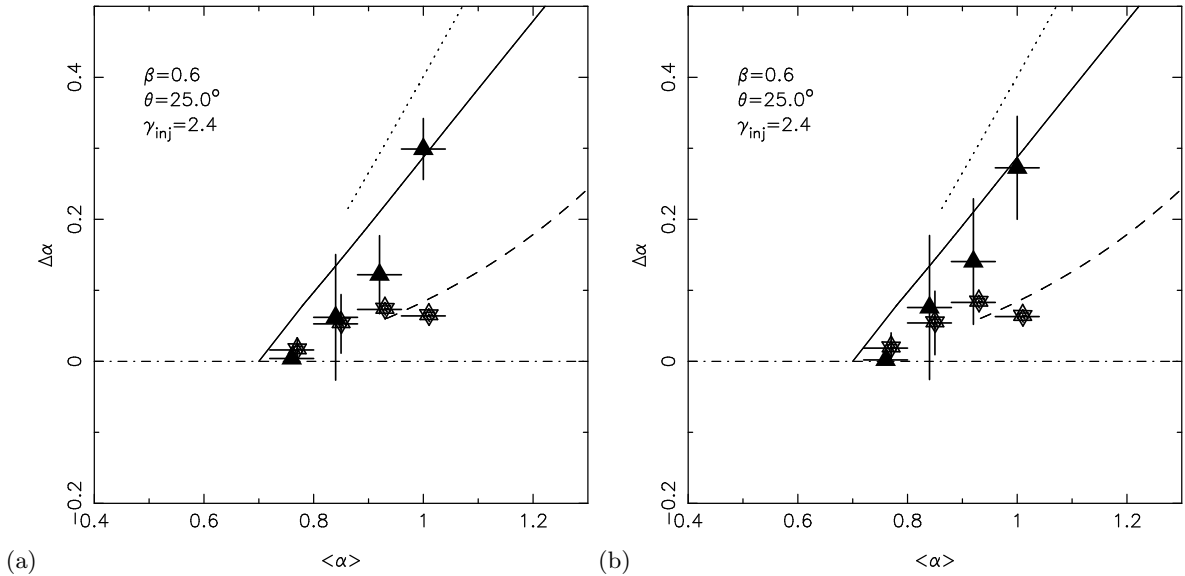

\centerline{(a)\psfig{file=h2439f6a.ps,angle=-90,height=7.5cm}
(b)\psfig{file=h2439f6b.ps,angle=-90,height=7.5cm}}
\caption{The spectral asymmetries of the BLRGs (stars) and quasars
(triangles) binned in mean spectral index of hotspots: (a) at original
resolution and (b) at similar linear resolution (see text).  The
dotted and dashed lines indicate the effect of addition of background
(lobe) material (model B, as before). Vertical error bars indicate spread
in the measurements; horizontal error bars indicate the binning interval.}
\label{fig:average}
\end{figure*}

 In this section we assess whether there is any evidence from the
spectral asymmetries for the BLRGs being statistically further from
the line of sight. We define 3C\,206, 3C\,246 and 3C\,323.1 as quasars
and the remainder of broad-lined objects in the sample observed for
this paper as BLRGs. We combine these with the BLRG and quasars from
D97 and D99, as in the previous sections. Binning the data in average
hotspot spectral index (Fig.~\ref{fig:average}), we see that the
quasars show a greater increase in $\Delta\alpha$ with mean $\alpha$
(3C\,17 has been excluded from this and following plots, and
accompanying analysis on the grounds of obviously insufficient
resolution.) From Fig.~\ref{fig:theory} we can see this could be
explained by smaller Doppler factors (larger angles to the line of
sight and/or smaller bulk jet speeds).

\begin{figure}
\psfig{file=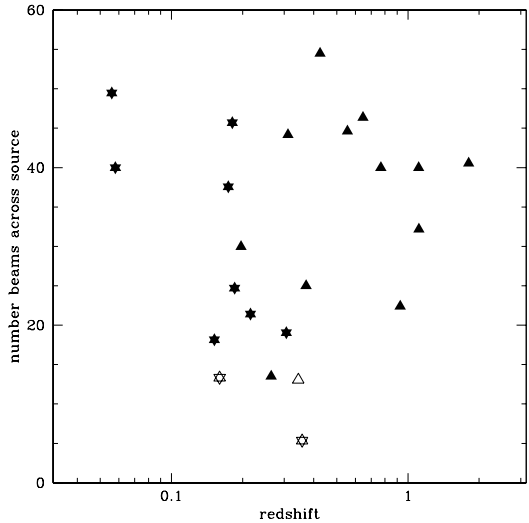,angle=0,height=7.5cm}
\caption{The number of beams across the sources in the observations
plotted in Fig.\ref{fig:average}(a).}
\label{fig:beam}
\end{figure}

However, as the observations stand, there is the possibility of more
lobe material in the `hotspot region' of the BLRGs.  We show in
Fig.~\ref{fig:beam} our linear resolution of the observed sources as a
function of redshift.  We note that there is a group of predominantly
BLRGs with $<$ 30 beams across the source, which could be preferentially
affected by lobe contamination.  We therefore have convolved all the
quasars and BLRGs of D97 and D99, and the BLRGs 3C\,33.1 and 3C\,219
from this paper to half the original resolution.  The final distribution
of linear resolution is then indistinguishable for the BLRGs and the
quasars.  From these images we derive the spectral asymmetry using the
peak of a Gaussian fit to the hotspot.  The results are plotted in
Fig.~\ref{fig:average}(b), and are very similar to those in
Fig.~\ref{fig:average}(a), showing that resolution effects are not the
cause of the difference. 

The BLRGs show a trend for smaller asymmetries at a given mean $\alpha$. 
This difference is suggestive that the average Doppler factor in the
hotspot material in the BLRGs may be smaller than the quasars.  It is
however, particularly dependent on the results from the BLRGs 3C\,109
and 3C\,234, which contribute the two BLRG points at high mean $\alpha$. 

We conclude that the observed spectral asymmetries are consistent with
the BLRGs having moderately relativistic jet speeds ($\beta=0.4-0.6$)
and angles to the line of sight preferentially directed towards the
observer.  There is an indication of smaller asymmetries in the BLRGs
which could be attributable to (some of the) the BLRGs in the sample
having lower Doppler factors (i.e.  larger angle to the line of sight,
or smaller bulk speeds). 

\subsection{Depolarisation asymmetries}

\begin{figure}
\psfig{file=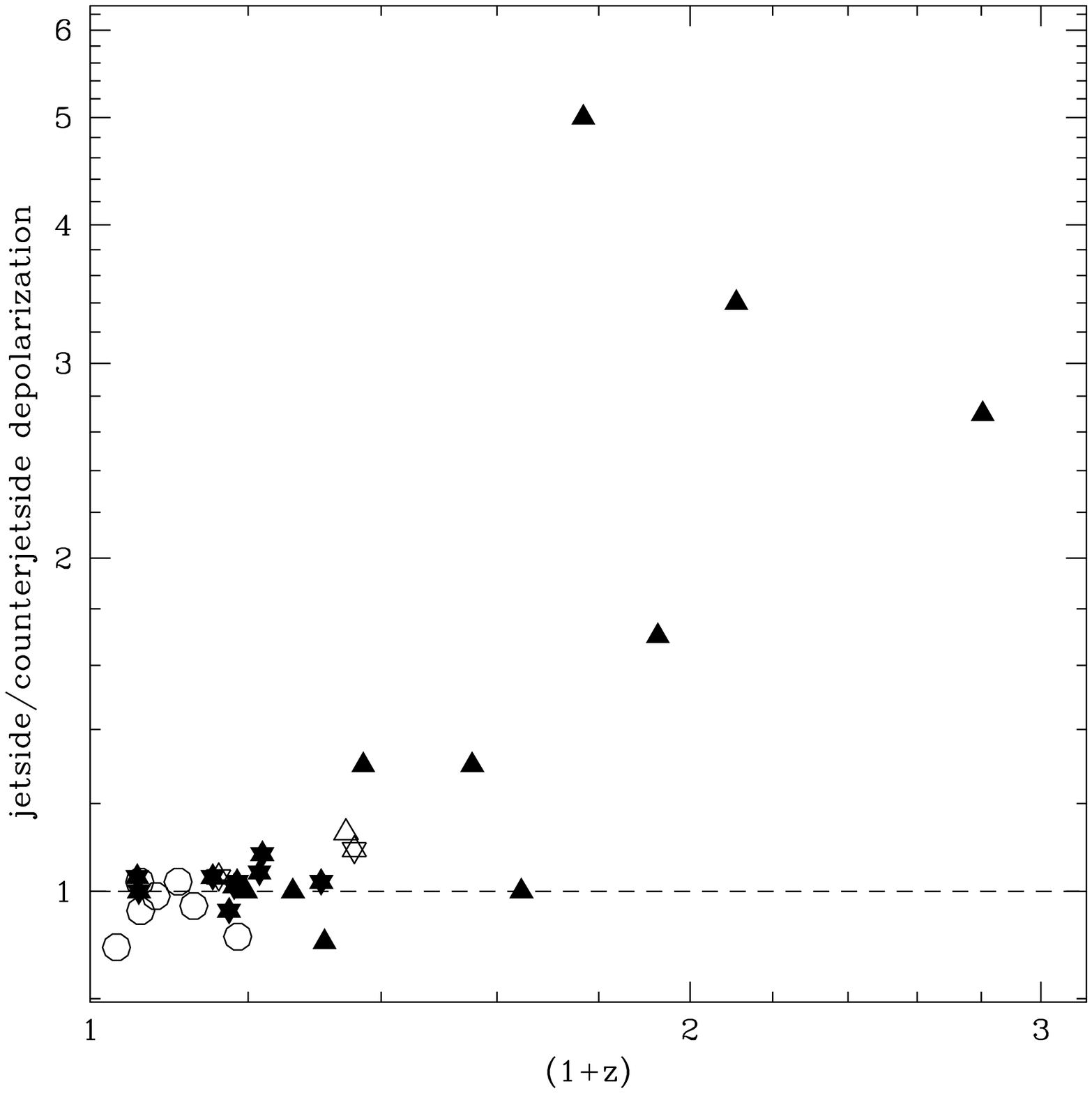,angle=0,height=7.5cm}
\caption{Depolarisation asymmetries of the objects in this sample,
together with data from D97 \& D99. Symbols as Fig. 1}
\label{fig:dp}
\end{figure}

The depolarisation asymmetries in the BLRGs, like the spectral
asymmetries, show the same sense but smaller magnitude 
compared to those in the quasars (D97)
(Fig.~\ref{fig:dp}).  Again, the smaller magnitude of the effect may be
attributed to power/ redshift as it is known that the depolarisation
of radio sources increases with power/redshift \cite{kro72,lea86a}.
Also, the magnitude of the asymmetry scales with strength of
depolarisation, both observationally, and as expected by theory if the
differential depolarisation is caused by differential lines of sight
though a surrounding magneto-ionic halo \cite{gar91a}, or superdisk
\cite{2000ApJ...529..189G}.  From the depolarisation properties we
also conclude that in general the BLRGs are at similar angles to the
line of sight as the quasars, but cannot rule out the possibility that
some individual cases may be at larger angles.

\begin{table*}
\caption{Optical source properties}
\label{tab:sources2}
\begin{tabular}{lllrrrl}
\hline
\hline
Name    & opt mag&A$_{\rm V}$& M$_{\rm tot}$&starlight & M$_{\rm AGN}$&ref\\
        &        &          &               & fraction &\\
\hline	       	     
3C\,17    & 18.02 &0.077  & --22.19 & 0.58 &    --21.25 & EH94 \\
3C\,33.1  & 19.5  &2.097  & --22.29 &  --  &     --     &      \\
3C\,93    & 18.09 &0.804  & --23.96 & 0.43 &    --23.35 & EH94 \\
3C\,109   & 17.88 &1.907  & --24.92 &$<$0.20&$<$--24.67 & M81  \\
3C\,206   & 16.5  &0.149  & --23.04 & 0.00 &    --23.04 & EH94 \\
3C\,219   & 17.22 &0.059  & --22.45 & --   &     --     &      \\
3C\,234   & 17.27 &0.062  & --22.54 & 0.35 &    --22.07 & T95  \\
          &       &       &         & 0.30 &    --22.15 & M81  \\%
3C\,246   & 16.00 &0.144  & --25.30 & 0.00 &    --25.30 & EH94 \\
3C\,287.1 & 18.27 &0.082  & --21.91 & 0.36 &    --21.42 & EH94 \\
          &       &       &         & 0.50 &    --21.16 & M81  \\
3C\,323.1 & 16.69 &0.140  & --24.00 &$<$0.1& $<$--23.89 & M81  \\
3C\,332   & 16.   &0.079  & --23.38 & 0.85 &    --21.32 & EH94 \\

\smallskip
3C\,381   & 17.46 &0.175  & -22.13 &  --  &    --     &      \\
3C\,382   & 14.73 &0.23   & -22.67 & 0.06 &    -22.60 & EH94 \\
3C\,390.3 & 14.37 &0.24   & -22.97 & 0.31 &    -22.57 & EH94 \\
\hline

\end{tabular}

\smallskip

{Optical magnitudes from Spinrad et al. (1985) for all sources
but 3C\,206 (a typical value from data presented in Ellingson et al.,
1989: the source shows $\pm$1\,mag variations) and 3C\,246. A$_{\rm V}$ from
NED. EH94=Eracleous \& Halpern 1994; M81=Miller 1981; T95=Tran
et al. 1995. 3C\,382 and 3C390.3 were added for comparison -- see
text}

\end{table*}

\section{Weaker or misdirected quasars?}
\label{sect:other}

As shown above, the scatter on the radio data means we are unable to
determine the line of sight to individual objects in this manner.  In an
attempt to distinguish between objects which may be misdirected, or
those which are intrinsically weaker with respect to their host
galaxies, we consider other available evidence. 
\nocite{1981PASP...93..681M}
\nocite{1983ApJ...269..352S}

In Table~\ref{tab:sources2} we present optical information from the
literature for the 12 broad-line objects in the redshift-limited
sample considered here. Although formally not members of the sample,
the low redshift ($z = 0.06$) BLRGs 3C\,382 and 3C\,390.3 are also
included.  Firstly consider the starlight fraction of the sources with
strong, unambiguous broad H$\alpha$ lines.  \cite[hereafter
EH94]{1994ApJS...90....1E} calculated the stellar contribution to
their high resolution spectra, and these have been used where
available as they represent by far the largest set of homogenous
estimates of this type.  However EH94's estimates are likely to be a
lower limit to the stellar contribution, because the total galaxy is
effectively under-represented in a slit across the nucleus and host.
Nonetheless, as we shall see, useful information can be derived,
despite the uncertainties.

We see that all objects classified as quasars in
Sect.~\ref{sect:class} have $<$10\% stellar light, essentially
confirming our independent classification.  Of interest is the optical
power of the AGN itself.  First we will consider the AGN power (i.e.
excluding the starlight) without adjusting for any intrinsic reddening
(but taking Galactic extinction into account following Cardelli et al.
1989).  The K-corrections are small and have been omitted.  The
resulting absolute magnitudes are listed in column 6 of
Table~\ref{tab:sources2}, and are denoted by M$_{\rm AGN}$.

Before considering any reddening of the spectrum associated with
absorption within the host galaxy, we can see that five sources (3C\,93,
3C\,109, 3C\,206, 3C\,246, 3C\,323.1) may be considered as `quasars'
from their optical luminosity alone (using Schmidt \& Green's
\cite*{1983ApJ...269..352S} cosmology adjusted M$<-22.4$ criterion). 
3C\,206, 3C\,246 and 3C\,323.1 were considered quasars by the criteria
in Sect.~\ref{sect:class}, but 3C\,93 and 3C\,109 were not.  As noted
in Appendix B, 3C\,109 is indeed an unusual object with apparently both
high intrinsic reddening and a very blue underlying continuum.  3C\,93
also has a highly reddened nucleus.  In contrast, there is no evidence
to suggest that 3C\,206, 3C\,246, 3C\,323.1 are anything other than low
redshift quasars: i.e.  at similar lines of sight and with similar
intrinsic powers as their higher redshift counterparts. 

The remaining seven broad-lined sources whose M$_{\rm AGN}$ is below
that for typical quasars include all of those whose lines show only
`broad wings'.  For all of these latter objects there is no
information about the percentage of the flux density due to starlight:
but these all have optical appearances of galaxies so we can safely
assume $<$ 50\% of the observed light comes from the AGN.  In this
case A$_{\rm V} \gtrsim$2 would be required to bring these objects
into the quasar luminosity range.  We suggest that all these objects
(if the broad wings on the spectral lines in 3C\,381 are confirmed) are
typically luminous quasars seen through the edge of the obscuring
torus.  X-ray observations and analysis are not available for all
sources in this sub-sample, but there is good evidence for excess
absorption towards 3C\,219 \cite{1999ApJ...526...60S}.  3C\,234 is
probably the best candidate for an object at intermediate line of
sight (see Appendix~B).

Finally we consider the three sources which have strong, unambiguous
broad lines, and are not the quasars discussed above (3C\,17, 3C\,332
and 3C\,287.1).  3C\,332 is a double peaked H$\alpha$ emitter (EH94). 
3C\,93 is the only other such object in the sample.  3C\,17 is also
classified by EH94 as a `disklike emitter' but has an asymmetric line
profile which cannot be fitted with a simple disk model.  It has a very
large H$\alpha$ FWHM (11500\,km/s), less than the FWHM of the double
peaked lines, but substantially more (at least a factor two) than all
other objects in this sample.  We do not fully understand how this class
of objects fits into the orientation model.  In order to become
typically luminous quasars these three objects are also required to
suffer reddening of their central continuum source (again typically
A$_{\rm V} \sim 2$).  3C\,287.1 may be either an intrinsically weak
quasar or seen through absorbing material: the available evidence is not
clear. 

Sambruna et al.  (1999) report a surprising amount of absorption due
to cold material in the ASCA spectra of {\it both} BLRGs and quasars.
They differentiate between BLRGs and quasars on the basis of [OIII]
luminosity.  [OIII] luminosity is related to the intrinsic power of
the source \cite{1989MNRAS.240..701R}, and is believed to be isotropic
to first order (Browne \& Murphy 1987; Jackson \& Browne 1991).
\nocite{bro87,jac91}  However, there is strong evidence that some
fraction of the [OIII] may be emitted in the obscured central regions
(Jackson \& Browne 1990; Hes et al. 1993).  Therefore,
while the division of Sambruna et al.  is primarily by intrinsic
power, it probably contains an additional orientation factor.  We wish
to divide the sources by orientation, so we reshuffle their objects
into categories determined by the suggested orientations in this
section.  When classified in this manner, with `BLRG' further
from the line of sight, it appears that the BLRGs in
the Sambruna et al.  sample have significantly higher absorption
column densities than the quasars, as expected.

In the light of these above considerations we divide the objects into
the following categories.\\
IA -- objects with strong broad lines, blue continuum, and a weak host 
3C\,206, 3C\,246, 3C\,323.1\\
IB -- objects with strong broad lines, a clear but  inconspicuous nucleus and
prominent host galaxy: 3C\,17, 3C\,332\\
IIA -- objects with  strong broad lines, a prominent, reddened nucleus,
and a relatively weak host: 
3C\,93, 3C\,109, 3C\,287.1\\
IIB -- radio galaxies with a broad winged component in their spectra:
3C\,33.1, 3C\,219, 3C\,234, 3C\,381\\

These classifications are based on observational parameters, but
divided with reference to simple, orientation dependent models as
follows.  Category IA contains the normal quasars, whilst less
luminous versions of these are found in IB.  The well-known nearby
BLRGs 3C\,382 and 3C\,390.3 are type I objects (both objects are
optically variable Type~1 Seyfert nuclei). Their AGN luminosity makes
them borderline IA/IB, but their low redshift makes their hosts
conspicuous. Their starlight fraction estimated by EH94 is probably an
underestimate, due to the effect of finite slit width. 

Class IIA objects are probably seen at grazing incidence to the
nuclear torus, with class IIB suffering more reddening and extinction
due to intervening (torus?)  dust.  The difference between classes IIA
and IIB may be a result of the optical AGN radiation of BLRGs being a
composite of dust scattered and dust transmitted light (Cohen et al.,
1999). Continued spectropolarimetry of BLRGs is crucial to address the
validity of our classification in the framework of the Cohen et al.
model.  Whether or not Class~II objects are at larger absolute angle
to the line of sight than the quasars is dependent on the opening
angle of the torus, which may exhibit large object to object
variations, or may vary systematically with source power
(e.g. Lawrence 1991; Hill et al. 1996).

\nocite{law91}  \nocite{1996ApJ...462..163H}

We note with interest that the smaller Doppler factor required to
explain the spectral asymmetries of the BLRGs compared to the quasars
rests on two class~II sources, namely 3C\,234 and 3C\,109.  As there
is good evidence that both these sources are observed at grazing
angles to their tori, their jet Lorentz factors may still be in the
range of objects in class IB (3C\,234) and class IA (3C\,109). In
summary, in agreement with Cohen et al. (1999), we consider it likely
that class II BLRGs are, {\it on average}, at larger angles than class
IA quasars.

\section{Conclusions}

Analysis of continuum radio images at two frequencies of a carefully
selected sample of broad lined radio galaxies and low redshift quasars,
reveals asymmetries in both depolarisation properties and spectra of the
hotspot regions.  These asymmetries are generally associated with the
side of the detected jet.  These results are in the same sense as the
asymmetries previously reported by Laing \cite*{lai88}, Garrington et
al.  \cite*{gar88}, Dennett-Thorpe et al.  (1997, 1999), and can be
understood as orientation effects: the depolarisation asymmetry as a
line of sight effect, the spectral asymmetry as due to relativistic
motion of the jet material on kiloparsec scales. 

The sense of the asymmetries are consistent with the previous results,
and reinforces the idea that the population of BLRGs is non-randomly
oriented and that they are at close angles to the line of sight, similar
to quasars.  The magnitude of both the spectral and the depolarisation
asymmetries is less than has been reported previously for the quasars. 
The smaller depolarisation asymmetry is explicable as a power/redshift
effect.  Sources at higher redshift in flux limited samples (i.e.  also
higher power) are known to depolarise more rapidly \cite{lea86a}, and
thus the asymmetries are more pronounced. 

The increasing spectral asymmetry with power/redshift may also be
easily explained if the asymmetry is caused by relativistic Doppler
effects in the hotspot material, if this has a curved spectrum. 
However, there is an indication of a difference in the magnitude of the
spectral asymmetry between the classes of BLRGs and quasars, which is
not a power/redshift effect, but could be explained by (some of) the
BLRGs in the sample having lower Doppler factors (larger angle, or
smaller bulk speeds) than the quasars.  Although this result depends on
just two sources, it is interesting to note that these two sources
(3C\,234, 3C\,109) are the ones for which there is good evidence from
other observations that they are viewed at grazing incidence to the
torus. 

As well as the problem of classification of objects by different workers
in different manners, there is the additional problem that both the
slightly `misdirected quasars' and the lower power objects at lines of
sight similar to the quasars are often classified, on observational
grounds, as BLRGs.  We have made comments on each individual source
regarding its likely position in a physically based orientation scheme
and consider that, whilst somewhat incomplete and uncertain, it is
preferable to most subjective or arbitrary numerical decisions that are
presently used. 

In summary, this work supports the idea the object class of BLRGs is
oriented preferentially towards our line of sight, and probably contains
objects at somewhat larger angles than the quasar population, as well as
less luminous quasars.  The warm dust peak, detected in some of the
BLRGs, has yet to be explained. 

\begin{acknowledgements}

We are grateful to the staff of the VLA for assistance, and a
rapid rescheduling of 3C\,286.  Thanks to Paul Alexander for the software
which produced the spectra used to calculate the asymmetries.  This
research has made use of the NASA/IPAC Extragalactic Database (NED)
which is operated by the Jet Propulsion Laboratory, California Institute
of Technology, under contract with the National Aeronautics and Space
Administration.  The National Radio Astronomy Observatory is a facility
of the National Science Foundation operated under cooperative agreement
by Associated Universities, Inc.  This research was supported by the
European Commission, TMR Programmme, Research Network Contract
ERBFMRXCT96-0034 `CERES'.

\end{acknowledgements}

\appendix
\section{Radio images at 5\,GHz}
\begin{figure*}
\hspace*{1.5cm} \raisebox{-1cm}{3C\,17} \hspace*{7.8cm} 3C\,33.1

\vspace{-1cm}

\centerline{\psfig{file=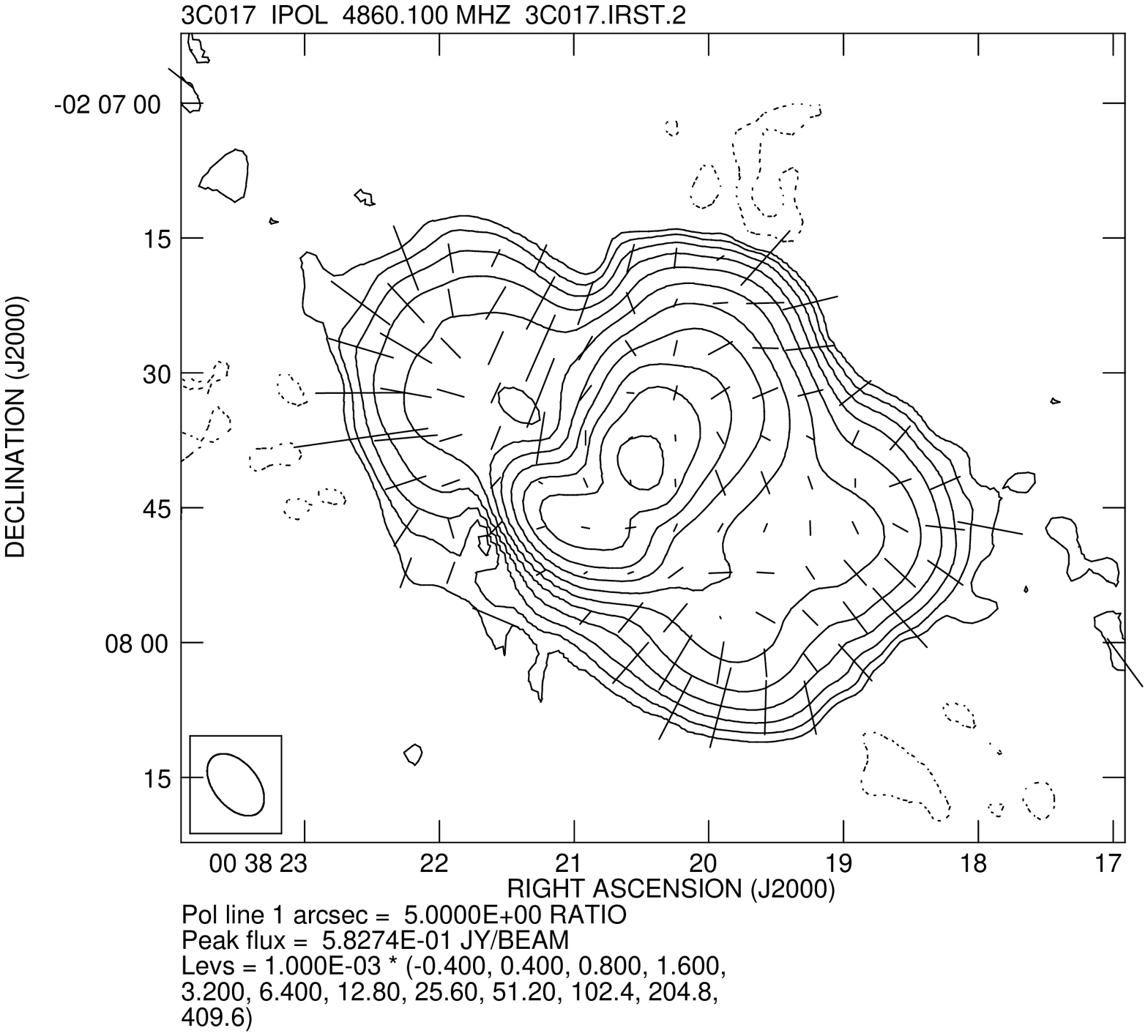,height=6.5cm,angle=-90,clip=}
\psfig{file=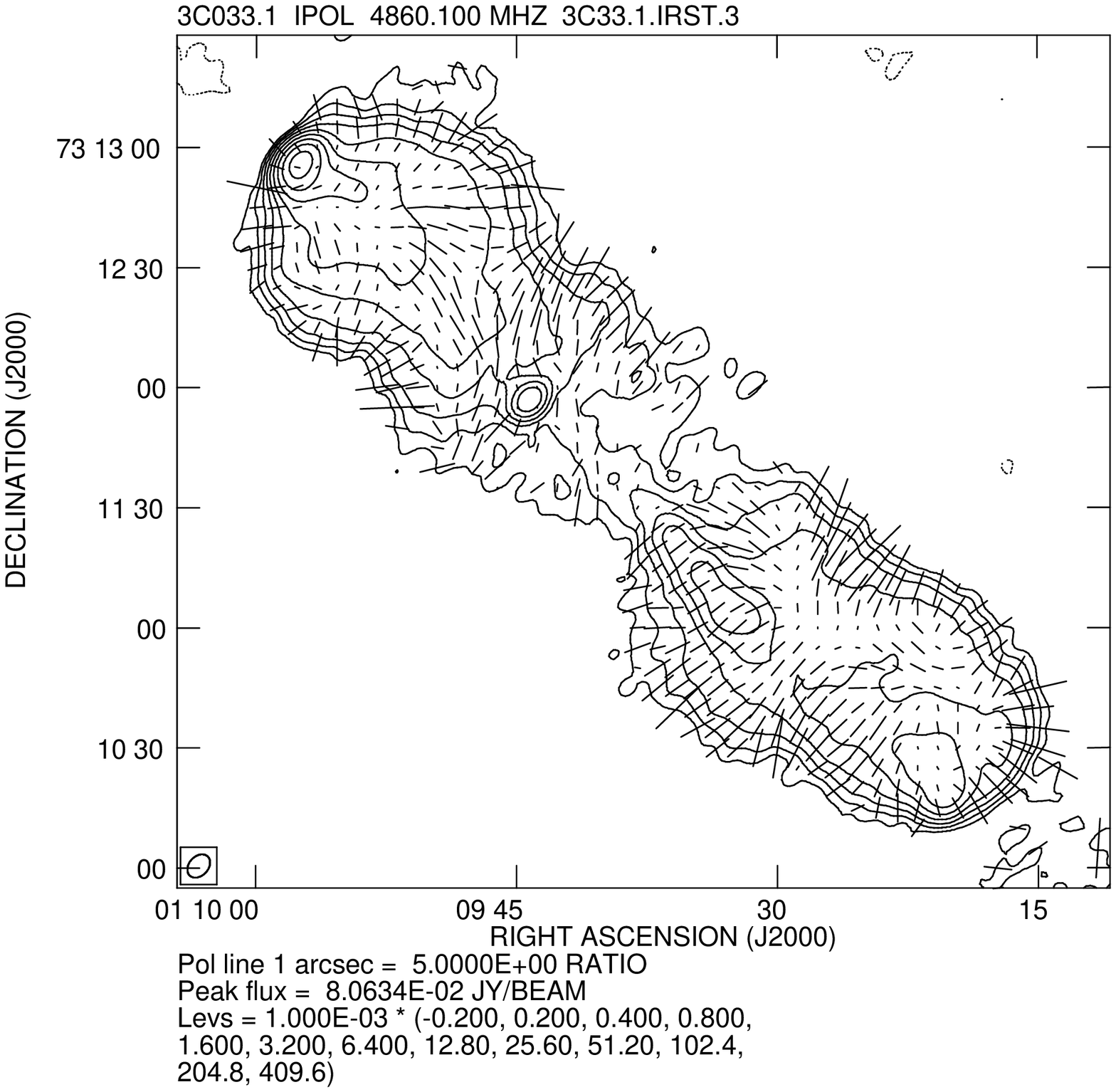,height=7.5cm,angle=-90,clip=}}

\hspace*{2.5cm} \raisebox{-.5cm}{3C\,93} \hspace*{7cm} 3C\,109

\vspace{-.5cm}

\centerline{\raisebox{.5cm}{\psfig{file=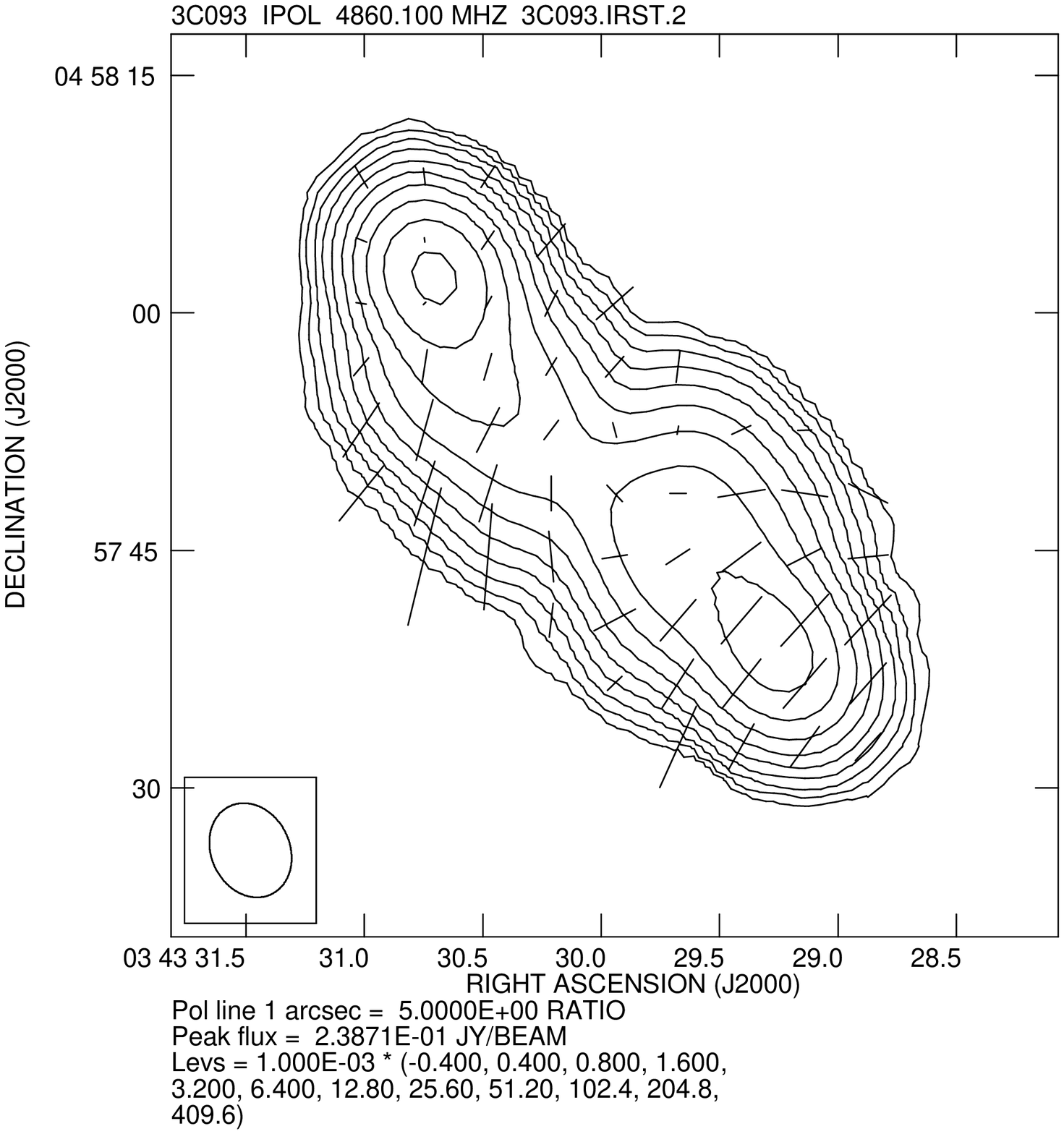,height=7.cm,clip=}}
\psfig{file=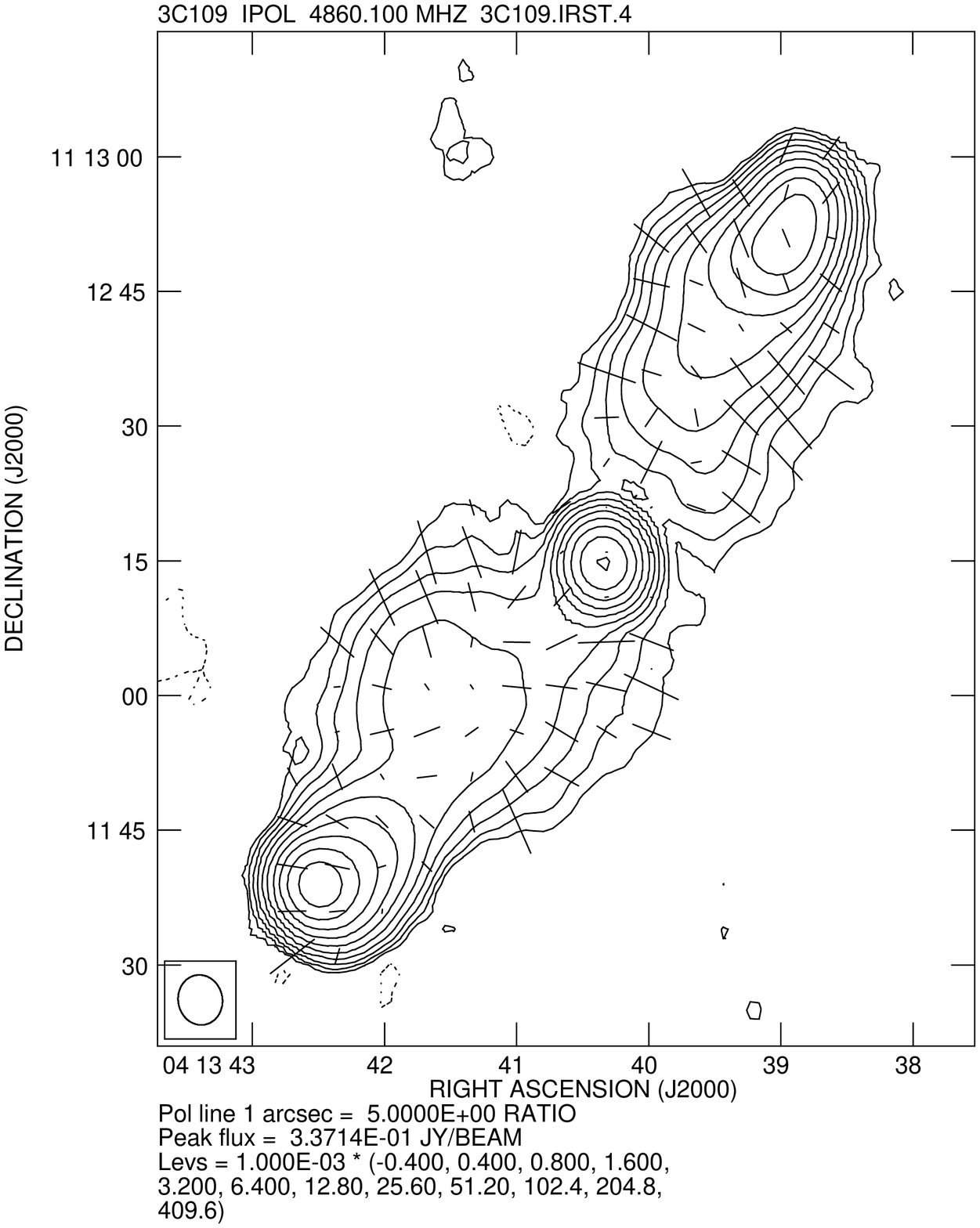,height=8cm,clip=}}

\hspace*{1.8cm} \raisebox{-.6cm}{3C\,206} \hspace*{8.8cm} 3C\,219

\vspace{-.6cm}

\centerline{\raisebox{1.cm}{\psfig{file=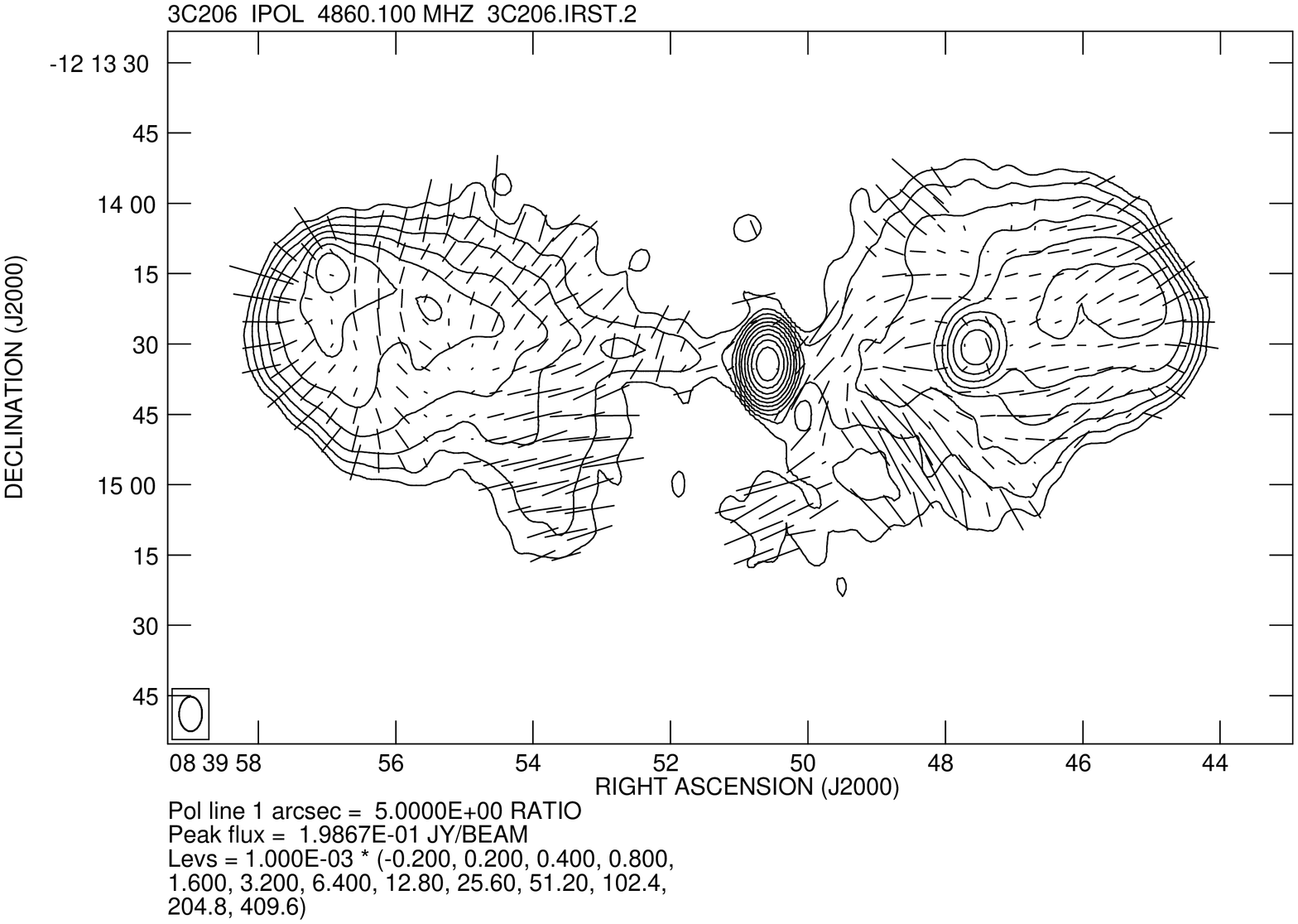,width=10cm,angle=-90,clip=}}
\psfig{file=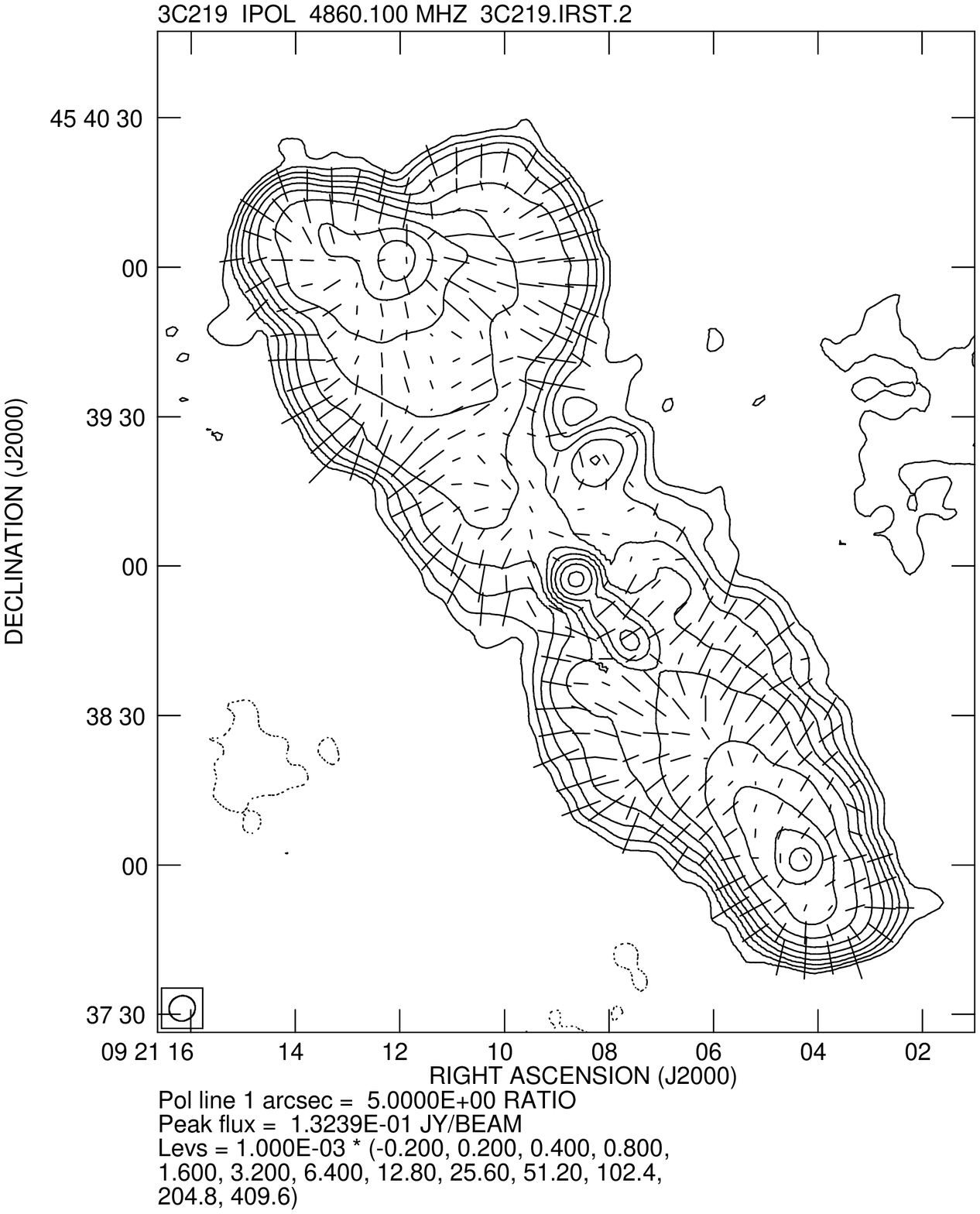,width=7.cm,clip=}}

\caption{5GHz (VLA in C array) images of the broad-lined objects in
the sample.  Contours are separated by factors of two in mJy/beam,
with lowest contours $\pm$ 0.2\,mJy/beam for 3C\,33.1,3C\,206,3C\,219
and $\pm$ 0.4\,mJy/beam for 3C\,17,3C\,93,3C\,109. Vectors represent
(RM uncorrected) apparent E-vectors with length proportional to
fractional polarisation, such that 1\,arcsec corresponds to 5\% polarisation.}

\end{figure*}
\begin{figure*}

\hspace*{2cm} 3C\,234 \hspace*{7.8cm} \raisebox{-.7cm}{3C\,246}

\vspace{-.7cm}

\centerline{\psfig{file=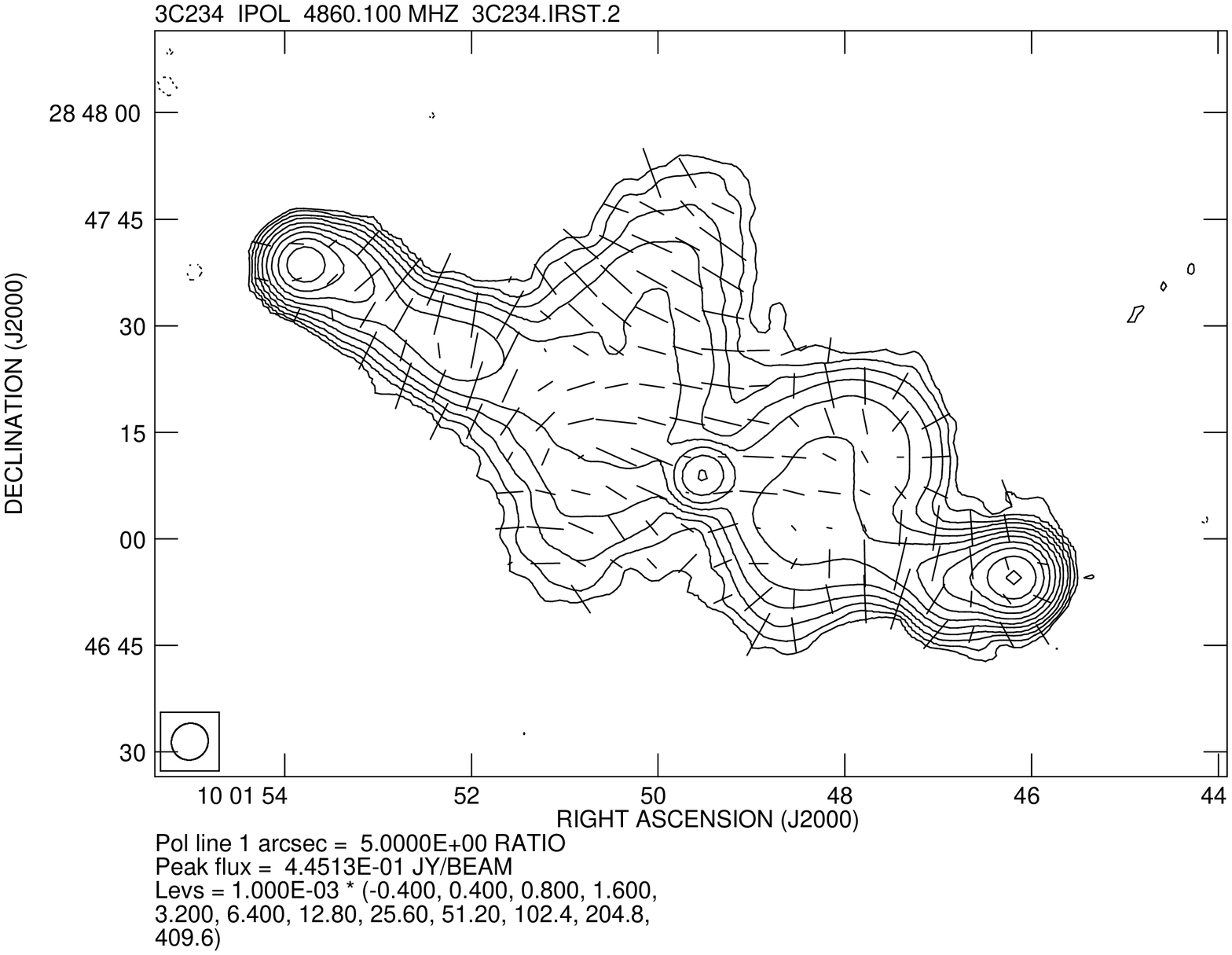,width=9cm,angle=-90,clip=}
\raisebox{.5cm}{\psfig{file=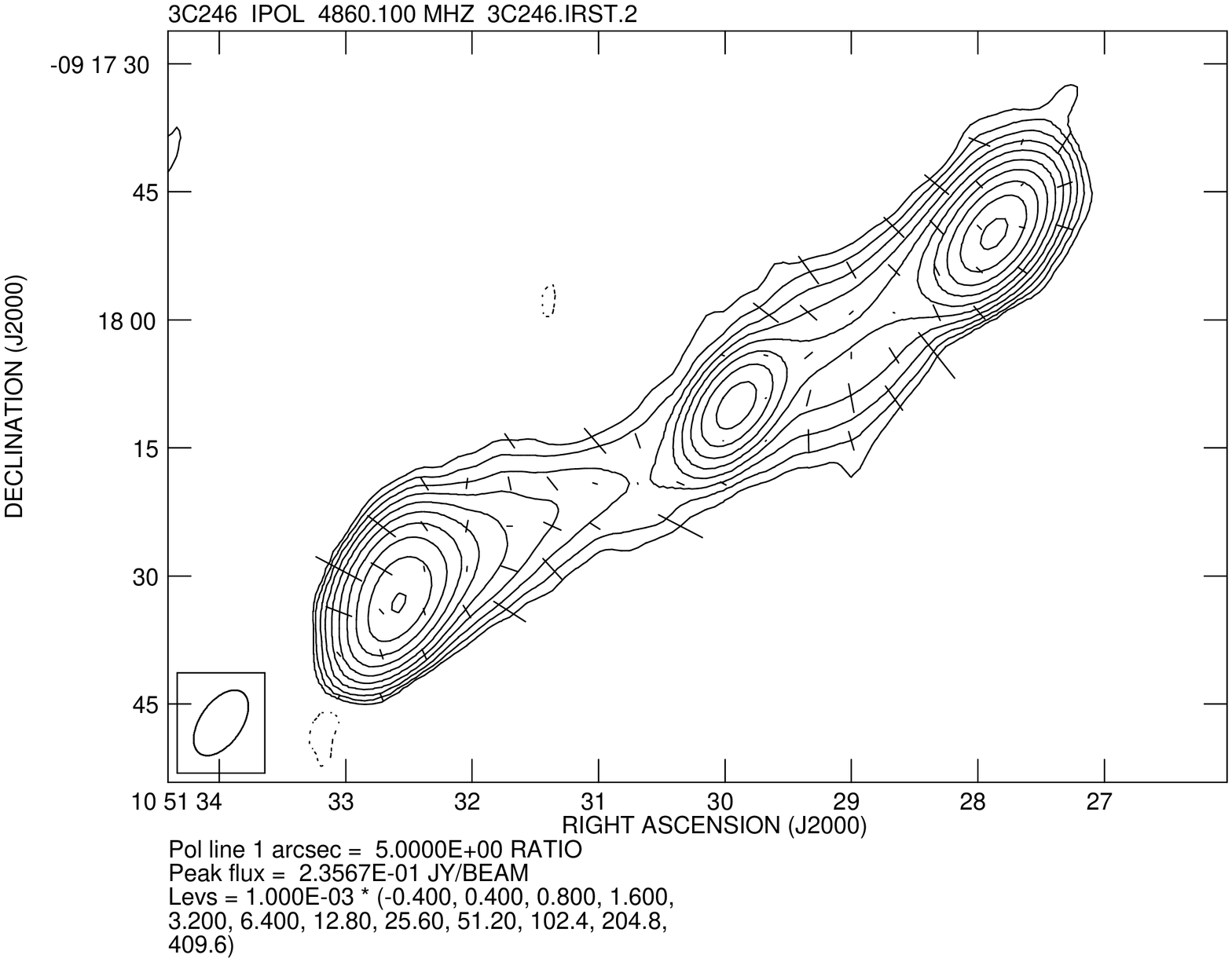,width=7cm,angle=-90,clip=}}}

\hspace*{2cm} \raisebox{-.5cm}{3C\,287.1} \hspace*{8.5cm} 3C\,323.1

\vspace{-.5cm}

\centerline{\raisebox{.7cm}{\psfig{file=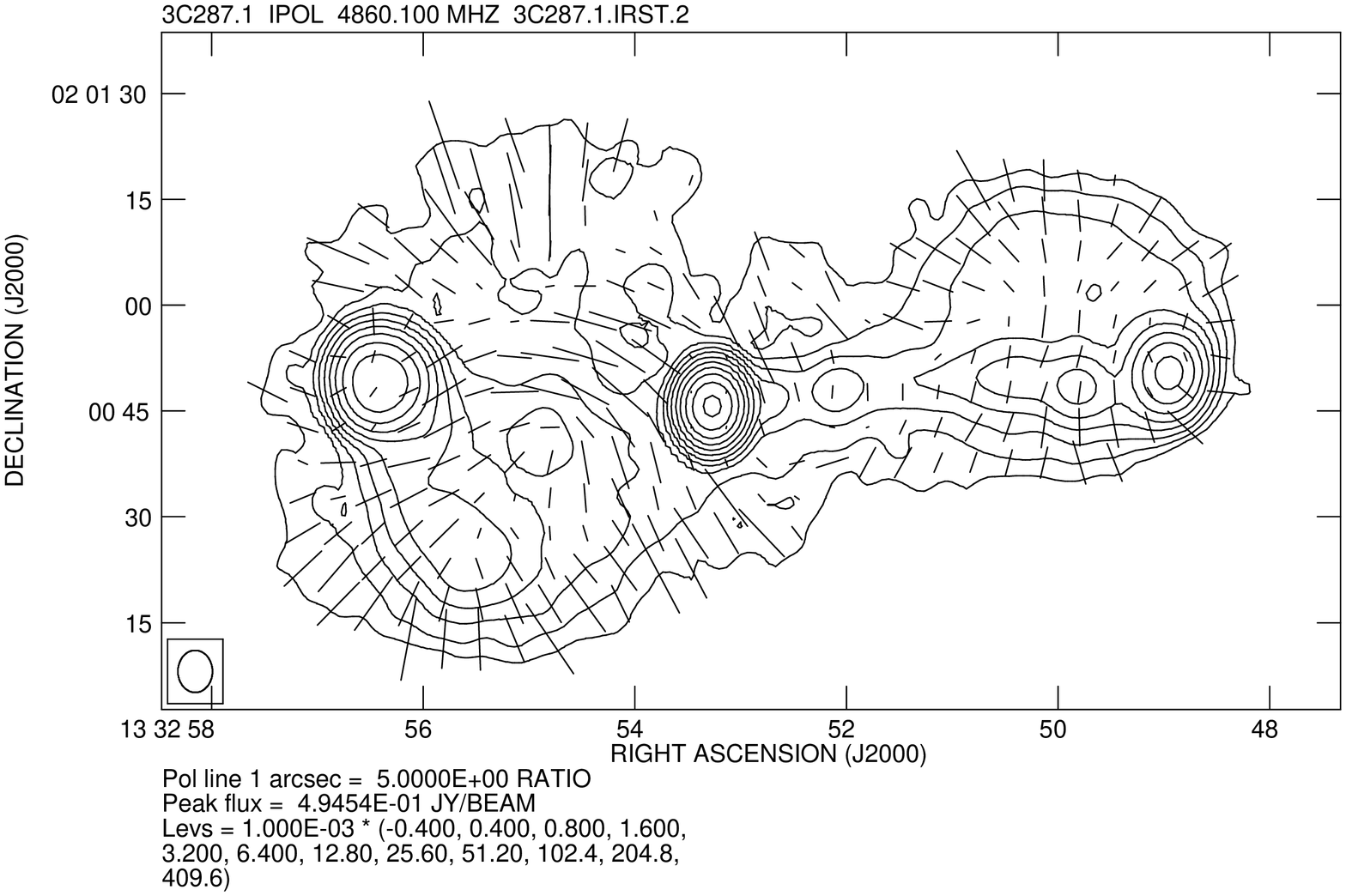,width=10cm,angle=-90,clip=}}
\psfig{file=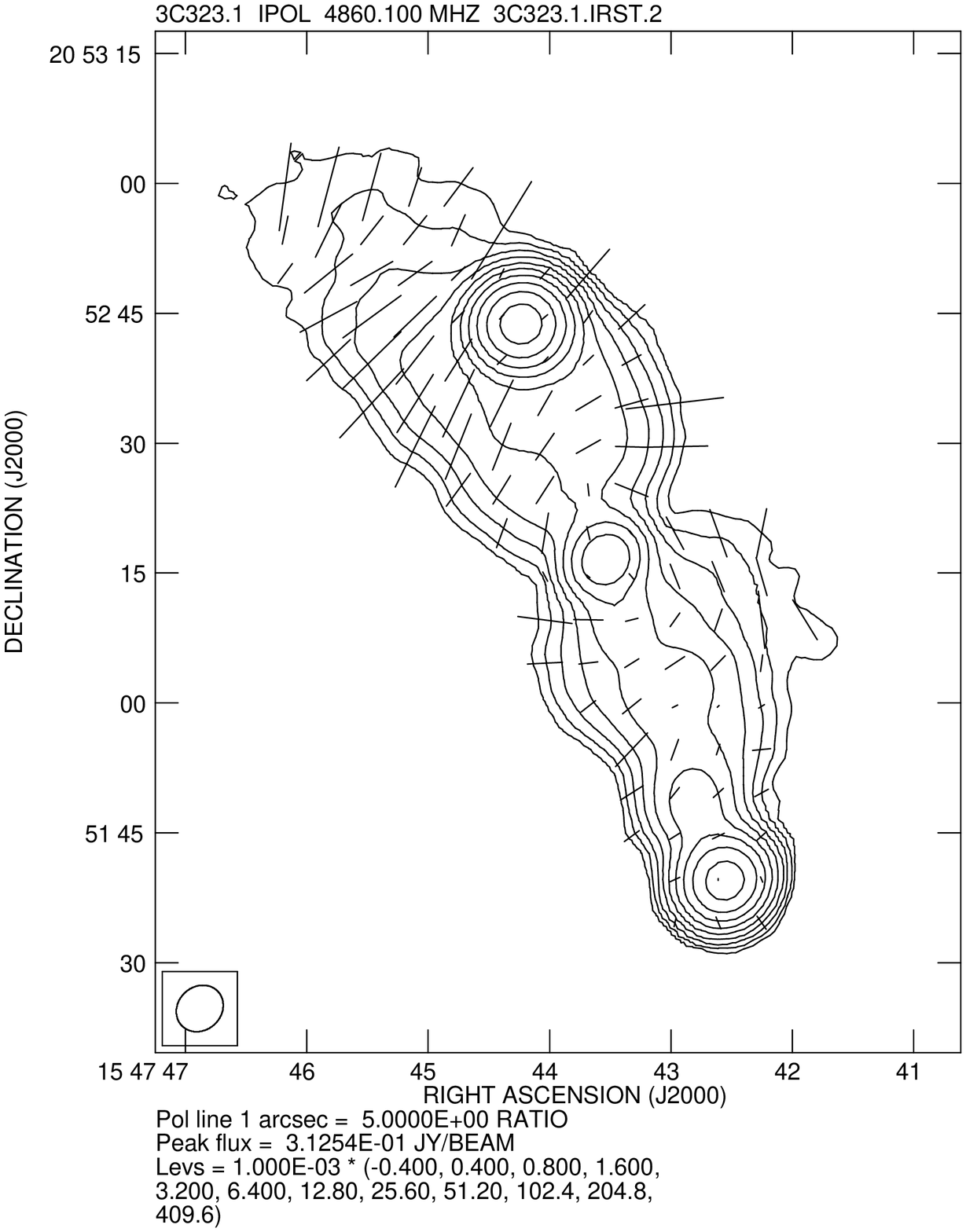,width=6cm,clip=}}

\hspace*{3.3cm} \raisebox{-.7cm}{3C\,332} \hspace*{6.cm} 3C\,381

\vspace{-.7cm}

\centerline{\raisebox{.5cm}{\psfig{file=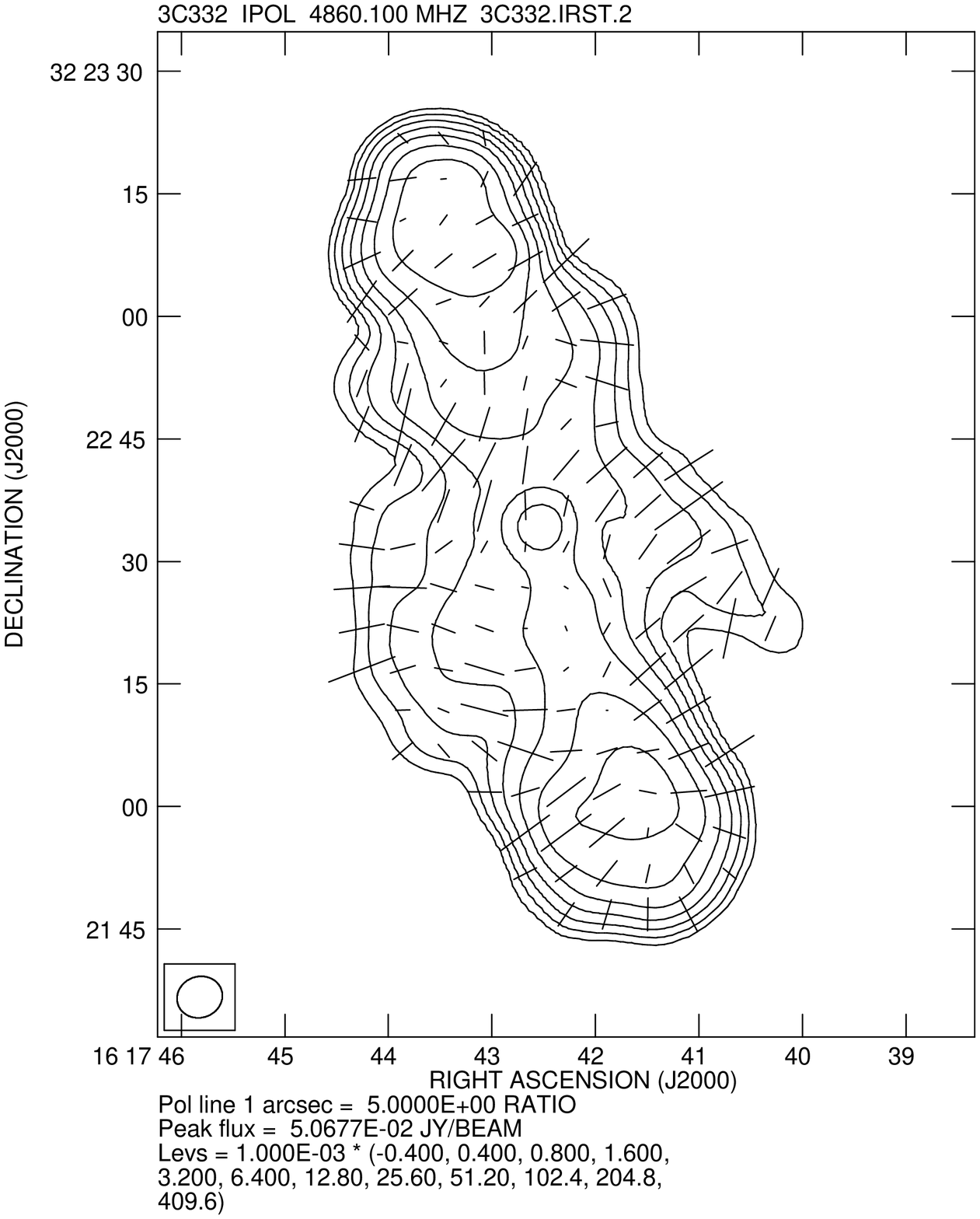,width=7cm,clip=}}
\psfig{file=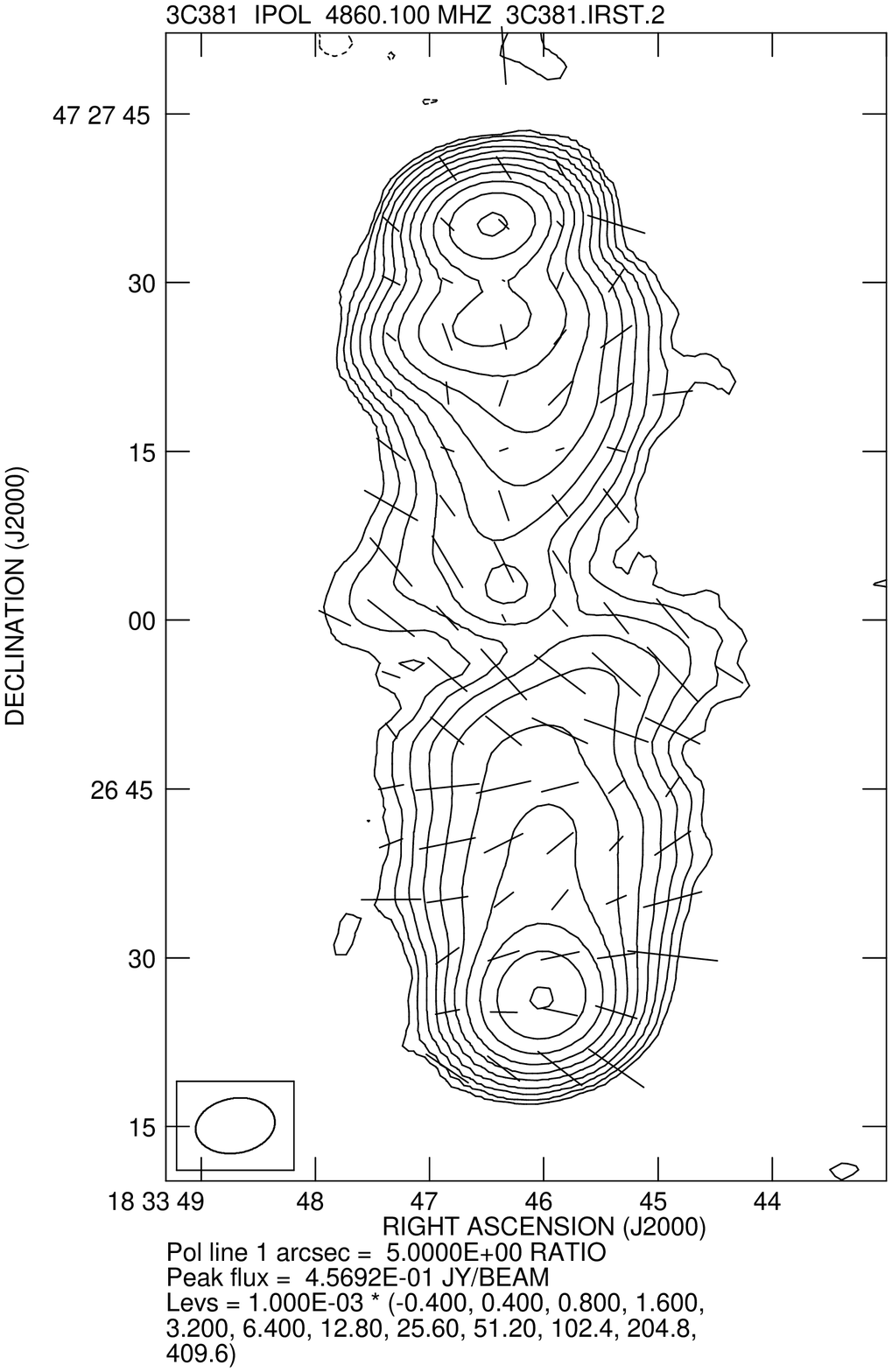,width=6.5cm,clip=}}
\end{figure*}

\begin{figure}
\psfig{file=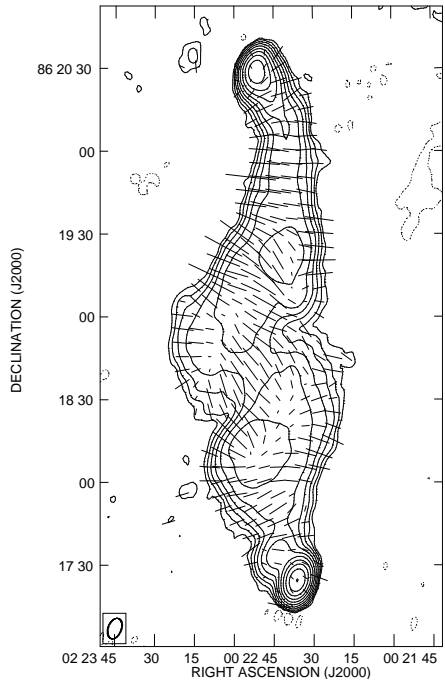,height=9cm,clip=}
\caption{5\,GHz image of the narrow-line radio galaxy 3C\,61.1 Contours are separated by factors of two in mJy/beam,
with lowest contours $\pm$ 0.4\,mJy/beam. Vectors as before.}
\end{figure}
\newpage

\section{Notes on individual sources}
\label{sect:notes}

\indent {\bf 3C\,17} (PKS\,0035$-$02) Morganti et al.
\cite{1999A&AS..140..355M} show this morphologically peculiar source
to contain a strong, bent jet on the south-eastern lobe.  The
H$\alpha$ line has a strong broad component (EH94), and the nucleus
shows optical polarisation \cite{1997quho.conf..311T}.  Venturi et al.
(2000) also find a VLBI jet in the same direction.  De Koff et al.
\cite*[hereafter K96]{1996ApJS..107..621D} show an HST image of the
host which shows a `compact nucleus'.

{\bf 3C\,33.1} The HST image (K96) shows a fairly undisturbed host
galaxy.  There is no published optical spectrum, but Laing et al. 
\cite*{1994pag..symp..201L} report broad wings on the H$\alpha$ line. 

{\bf 3C\,61.1} is a narrow-line galaxy, and was included erroneously in
our original list.  Lawrence et al.  \cite*{1996ApJS..107..541L} give a
narrow-line spectrum for the western object in the triplet of galaxies. 
Giovanni et al.  \cite*{1988A&A...199...73G} give a position for a radio
core which falls on this galaxy.  The BSO at the same redshift,
identified by Penston (1971) and Miller, Robinson \& Wampler
\cite*{1973ApJ...179L..83M} falls outside the radio contours (Longair \&
Gunn \cite*{1975MNRAS.170..121L} give the finding charts).  This source
has lead to some confusion with 3C\,61.1 appearing in lists of
broad-lined objects.  K96 note tails in the HST image of the host. 

{\bf 3C\,93} This source has been classified as a BLLac due to the weak
emission lines in the spectrum of Smith \& Spinrad (1980), although it
is now usually considered one of the nearest 3C quasars.  It has very
broad lines and is red (EH94).  Hes et al. \cite*{1996A&A...313..423H}
classify it as a BLRG due to its red colour and the slightly fuzzy R
band continuum.  EH94 derive a line of sight at 45$\pm$15$^\circ$, based
on disk models for the double-peaked H$\alpha$ line profile. 

{\bf 3C\,109} The nucleus is reddened (A$_{\rm V} \sim 2.7$)
\cite{1984ApJ...278..530R,1992ApJ...391..623G,1999AJ....118..666R}. 
De-reddening the nuclear source turns it into a typically luminous
quasar \cite{1992ApJ...391..623G}.  Obscuring material is seen as excess
absorption in the X-ray \cite{1992MNRAS.258P..29A,1997MNRAS.286..765A}. 
This source falls outside the simple two-component model for the optical
emission (direct light plus scattered polarised light) of Cohen et al. 
\cite*{1999AJ...118.1963C}: it requires either excess absorption towards
the BLR or an extra blue component in the continuum.  Modelling the
broad X-ray Fe line as emitted in the accretion disk, Allen et al. 
\cite*{1997MNRAS.286..765A} derive an inclination of $>35^\circ$.  The
authors suggest that this object is at intermediate angle to the line of
sight, grazing the torus.  Giovanini et al. \cite*{1994ApJ...435..116G}
used the core/jet and jet/counterjet flux ratios to derive an angle to
the line of sight of $34^\circ < \theta <56^\circ$.  A very compact
nucleus and tails are seen on the HST images (K96).  Taking the inferred
intrinsic reddening into account Goodrich \& Cohen (1992) determine
M$_{\rm V} < -26.6$

{\bf 3C\,206} is a fairly well-studied quasar.  It is a member of a rich
cluster \cite{1989AJ.....97.1539E}, and shows optical variations of up
to 2\,~mag in B.  Ellingson et al.  decompose the object into its host
galaxy and non-thermal component and find: V=18.45; B$-$V=1.25 for host
galaxy. 

{\bf 3C\,219} displays narrow lines in the optical with broad wings on
the H$\alpha$ \cite*{1994pag..symp..201L}.  Fabbiano et al.
\cite*{1986ApJ...304L..37F} report a strong Paschen-$\alpha$ line in
the infra-red, and suggested this line was broad, a claim confirmed by
Hill et al. \cite*{1996ApJ...462..163H}, who measured
A$_{\rm V}$=1.8$\pm$0.3.  HST observations show a compact nucleus
(K96).  A strong jet is detected to the south
\cite{1980AJ.....85..499P,1992ApJ...385..173C}.

{\bf 3C\,234} shows both narrow and broad components in the H$\alpha$
line, but only narrow components in the H$\beta$ and higher Balmer lines
\cite{1979ApJ...232..659G}.  Some authors therefore refer to it as
`narrow line' \cite{1996MNRAS.283..930T} and others `broad line'
\cite{1984ApJ...284..523C}.  Tran et al.  \cite*{1995AJ....110.2597T}
show that most of the broad line flux is due to the polarised emission,
with a very high (25\%) intrinsic polarisation, and argue for scattering
as the cause of the polarisation.  It is optically variable
\cite{1981PASP...93..681M}.  The BLR suffers heavy reddening, with an
A$_{\rm V}$ calculated from Pa$\alpha$/H$\alpha$/H$\beta$ line ratios of
$\sim$2.4 \cite{1984ApJ...284..523C,1996ApJ...462..163H}.  High column
densities are also detected in the X-ray \cite{1999ApJ...526...60S}. 
3C\,234 would therefore seem, like 3C\,109, to be a good candidate for
an object at intermediate angles to the line of sight.  Tails are seen
in the HST image (K96).  Applying corrections for reddening and
starlight contamination, Tran et al.  (1995) derive M$_{\rm
V}\lesssim-$24.2, i.e.  well within the range of typical quasars. 

{\bf 3C\,246} (PKS\,1048$-$090, PG1048$-$090) This object is also a member
of the Bright Quasar Survey
\cite{1983ApJ...269..352S,1986ApJS...61..305G}, so has been extensively
studied at optical wavelengths.  It is also a member of a cluster
\cite{1989MNRAS.240..129Y}.  No jet has yet been detected, but the
source is not a member of the LRL sample, and has been relatively poorly
imaged in the radio, in comparison with many other of the sources
\cite{1998PASP..110..111H,1994AJ....108.1163K}. 

{\bf 3C\,287.1} The HST image shows a compact nucleus, as well as a
second compact region to the W, less than 1kpc away (K96).  A strong jet
to the west, first indicated in Antonucci \cite*{1985ApJS...59..499A},
confirmed in Harvanek \& Hardcastle \cite*{1998ApJS..119...25H}, is also
clearly seen in our images.  Analysis of the RASS data shows 3C\,287.1
was detected with a photon index of $\approx$ 1.86
\cite{1998MNRAS.301..261S}, which is a value in the `overlap region'
between quasars and radio galaxies.  Crawford \& Fabian
\cite*{1995MNRAS.273..827C} show the source has excess soft X-ray
absorption. 

{\bf 3C\,323.1} (PG1545+210) This has been a target for imaging
investigations of the host galaxies of quasars (e.g.  Stockton 1982). 
Miller \cite*{1981PASP...93..681M} concluded that the host galaxy must
contribute $<$ 10\% of the optical light.  Imaging with the HST has now
detected its `low surface brightness elliptical'
\cite{1997ApJ...479..642B}.  Boroson \& Oke \cite*{1984ApJ...281..535B}
showed that the `nebulosity' surrounding the quasar is highly ionised. 

{\bf 3C\,332} The HST image shows a `smooth elliptical nuclear region'
(K96).  It appears to be a member of a close pair.  Halpern
\cite*{1990ApJ...365L..51H} suggests this object as the prototype for
relativistic accretion disk emission in the double-peaked H$\alpha$
line, and EH94 derive a line of sight of 36$\pm$3$^\circ$ based on
fitting disk models to the profile of this line. 

{\bf 3C\,381} A compact nucleus and tails appear in the HST image (K96). 
Grandi \& Osterbrock \cite*{1978ApJ...220..783G} claim `a broad weak
component of H$\alpha$', but to our knowledge there is no published
spectrum of H$\alpha$ to date.  We suggest that its identification as a
genuine BLRG may be open to doubt: it has no detected radio jet, and,
unlike all other BLRG observed by Lilly \& Longair
\cite*{1982MNRAS.199.1053L} its K band flux density is not higher than
the NLRG in the sample.

\end{document}